\documentclass[12pt]{article}

\usepackage{amssymb}
\usepackage{euscript} 
\def\ln{{\rm ln}}
\def\a{\begin{eqnarray}}
\def\b{\end{eqnarray}}
\def\0{\nonumber}
\def\ba{\begin{array}}
\def\ea{\end{array}}
\def\noal{\noalign{\vskip10pt}}

\newcommand\ED{\EuScript{D}}
\def\um{\frac{1}{2}}
\def\A{{\cal A}}

\def\Tr{{\rm Tr}}
\def\z{{\bf z}}

\newcommand{\ii}{\mathrm{i}}
\newcommand{\de}{\mathrm{d}}

\def\P{{\hat P}}
\def\q{{\widetilde{\cal Q}}}

\def\e{\epsilon}

\def\al{{\alpha}}
\def\lm{{\lambda}}

\def\cm{{\cal M}}

\def\tl{{\widetilde L}}
\renewcommand{\theequation}{\thesection.\arabic{equation}}

\newlength{\extraspace}
\newlength{\extraspaces}
\newcounter{dummy}
\newcommand{\ai}{
\addtocounter{equation}{1}
\setcounter{dummy}{\value{equation}}
\setcounter{equation}{0}
\renewcommand{\theequation}{\thesection.\arabic{dummy}\alph{equation}}
\begin{eqnarray}
\addtolength{\abovedisplayskip}{\extraspaces}
\addtolength{\belowdisplayskip}{\extraspaces}
\addtolength{\abovedisplayshortskip}{\extraspace}
\addtolength{\belowdisplayshortskip}{\extraspace}}
\newcommand{\bj}{
\end{eqnarray}
\setcounter{equation}{\value{dummy}}
\renewcommand{\theequation}{\thesection.\arabic{equation}}}
\def\d{{\partial}}

\newcommand{\ddlm}[1]{{\partial \over \partial \lm_{#1}}}

\def\ddt2{{{\d}\over{\d t_2}}}

\newcommand{\bac}{\begin{array}{c}}
\newcommand{\bacc}{\begin{array}{cc}}
\newcommand{\baccc}{\begin{array}{ccc}}
\newcommand{\barcl}{\begin{array}{rcl}}
\newcommand{\bacccc}{\begin{array}{cccc}}
\newcommand{\baccccc}{\begin{array}{ccccc}}
\newcommand{\baccccccc}{\begin{array}{ccccccc}}
\newcommand{\barclcrcl}{\begin{array}{rclcrcl}}
\newcommand{\bacl}{\begin{array}{cl}}
\newcommand{\bal}{\begin{array}{l}}
\newcommand{\bacll}{\begin{array}{cll}}
\begin{document}
\begin{flushright}
SISSA-ISAS 55/2005/EP\\
hep-th/0507224
\end{flushright}
\vskip0.5cm
\centerline{\LARGE\bf Conifold geometries, topological strings}
\vskip0.5cm
\centerline{\LARGE\bf  and multi--matrix models}
\vskip 1cm
\centerline{\large G.Bonelli, L.Bonora, A.Ricco}
\centerline{International School for Advanced Studies (SISSA/ISAS)}
\centerline{Via Beirut 2, 34014 Trieste, Italy}
\centerline{INFN, Sezione di Trieste.  }
\vskip5cm
\abstract{We study open B-model representing D-branes on 2-cycles of local 
Calabi--Yau geometries. To this end 
we work out a reduction technique linking D-branes partition functions
and multi-matrix models in the case of conifold geometries so that
the matrix potential is related to the complex moduli of the conifold.
We study the geometric engineering of the multi-matrix models and focus on
two-matrix models with bilinear couplings. We show how to solve this models
in an exact way, without resorting to the customary saddle point/large N 
approximation. The method consists of solving the quantum equations of motion
and using the flow equations of the underlying integrable hierarchy to derive
explicit expressions for correlators. Finally we show how to incorporate in this
formalism the description of several group of D-branes wrapped around
different cycles.}

\vfill\eject

\section{Introduction}

Singular Calabi--Yau spaces are more and more frequently met in 
string compactifications. The reason is mostly the fact that compactifications
on regular Calabi-Yau spaces do not seem to be able to describe crucial 
features of realistic physics. On the contrary the presence of a 
conifold point, \cite{candelas}, in a Calabi-Yau opens new prospects: in 
conjunction with fluxes
and branes it may allow for warped compactifications, which in turn may create
the conditions for large hierarchies of physical scales. On the other hand
singular Calabi-Yau's with conifold singularities seem to be 
necessary in order to realize low energy theory models with realistic 
cosmological features.
The hallmark of a conifold is the possibility 
of resolving the conifold point in two different ways, by a 2-sphere
(resolution) or a 3--sphere (deformation). This leads, from a physical point 
of view, to a geometric transition that establishes a duality relation 
between the theories defined by the two nonsingular geometries (gauge--gravity 
or open--closed string duality),\cite{Cachazo1,Cachazo2}. In summary conifold 
singularities are at the 
crossroads of many interesting recent developments in string theory. It has 
therefore become customary to study theories defined on conifolds, i.e. 
singular non compact Calabi-Yau threefolds, as calculable and well--defined 
models to approximate more realistic situations.    

Given the crucial role they play it is of upmost importance to find methods of 
calculation for theories defined on conifold geometries. In this sense
two main tools have been devised: topological field theories and matrix 
models. Topological field theories are truncations of full theories: one gives
up the knowledge of the dynamical sectors of a given theory, drastically 
simplifies it by limiting it to the topological sector and ends up with a 
theory where very often many quantities (correlators) can be explicitly 
calculated. However even topological field theories are sometimes not 
easily accessible to explicit calculations. Here come matrix models to 
the rescue. Sometimes, like in the examples of this paper, topological field 
theories can be shown to be equivalent to matrix models. This makes life 
easier, especially when the matrix models have couplings of special type.
In this case one can rely on the integrable structure
underlying the model (the Toda lattice hierarchy, \cite{UT}) which usually provides 
algorithmic methods to obtain the desired results in an exact and controlled
framework. The case of matrix models 
with more general couplings is more complex and represents a challenge that 
people have started to tackle only very recently.

In this paper we would like to elaborate on an idea that has recently received 
increasing attention: how data about the geometry of a local
Calabi--Yau can be encoded, via a topological field theory, in a (multi--)matrix 
model and how they can be efficiently calculated. The framework we consider is
IIB string theory with spacetime filling D5--branes wrapped around 
two--dimensional cycles. This geometry
defines a 4D gauge theory, \cite{KW,Dijk1,Dijk2,Dijk3}. On the other hand we can consider 
the open topological B model
representing the strings on the conifold. The latter has been 
shown by Witten long ago to be represented by a six-dimensional holomorphic 
Chern--Simons theory, \cite{Witten}. When reduced to the two-dimensional cycle 
this theory 
can be shown to boil down to a matrix model. In particular, if we wish 
to represent the most general deformations of the complex structure 
satisfying the Calabi--Yau condition, we end up with very general 
multi--matrix models. This point of view was advocated in \cite{Ferrari}. 
In this paper we concentrate on the topological string theory
part of the story, and ignore both the 4D gauge theory part and the closed
string theory side, which is attained by shrinking the resolved sphere to a
point and passing to the deformed picture in which the singularity is replaced
by a three--sphere (as would be possible at least for the cases corresponding to
one and two--matrix models).

As already pointed out the general idea underlying our paper has already been
developed in a number of papers, \cite{agan,Diaconescu,Marino,Bilal,neitzke}. 
Here we would like to concentrate
on particular aspects that have not been stressed or have been left aside in 
the previous literature. The first question we concentrate on is the reduction 
from the six-dimensional holomorphic Chern--Simons theory to a two--dimensional 
field theory. We wish to understand what degree of arbitrariness this passage 
implies, so as to be able to assess whether the information we gather from the
reduced theory is intrinsic or depends on the reduction process. Our conclusion 
is that the reduced theory does not depend on the reduction procedure. 

The second point we deal with is whether there are limitations on the general 
form of the potential we find for the multi-matrix model. We do find some 
conditions although rather mild ones. Finally we concentrate on the subclass of 
matrix models represented by two--matrix models with bilinear coupling. In this 
case the functional integral can be explicitly carried out with the method of 
orthogonal polynomials. We show, using old results, how one can find explicit
solutions: this is done by solving the {\it quantum equations of motion} and 
utilizing the recursiveness guaranteed by integrability. All the data turn
out to be encoded in a Riemann surface (plane curve), which we call 
{\it quantum Riemann surface} in order to distinguish it from the Riemann 
surface of the standard saddle point approach.  In particular we are able 
to prove that the exact solutions found in this way are {\it more} in number 
than the ones found by the saddle point method.  

The paper is organized as follows. 
In the next section we calculate the reduction of the
B-model open string field theory corresponding to the wrapped D-branes
on the 2-cycle of the conifolds and how it depends on its complex moduli.
The result is given in terms of a family of multi-matrix models.
In section 3 we discuss explicitly what multi-matrix models we do get
and we draw the geometrical engineering scheme for their realization
with D-branes. In particular we show how to obtain two-matrix models 
with bilinear couplings. In section 4 we review some general properties 
of the above class of 2-matrix models, their
integrability and the relative genus expansion.  
The section is concerned with particular method of solving them which is
based on the 
quantum equations of motion and the flow equations. Section 5 contains
several explicit examples of models solved with this method. 
We compare these solutions with the ones obtained with the usual saddle 
point/large N expansion. All these solutions are interpreted as describing
the physics of $N$ D-branes wrapped around a 2-cycle.
In section 6 we show how the physics of several group of D-branes wrapped 
around different cycles can be incorporated in the exact scheme proposed 
in this paper. Finally, section 7 is devoted to some conclusions and open 
questions. The Appendix extends the approach of Section 2 to local CY 
geometries around 4 cycles.

\section{Reduction to the branes and multi-matrix models}

In this section we show in detail how the reduction of the topological open 
string field theory (B model) to a 2-cycle in a local CY geometry is equivalent
to a multi-matrix model whose potential is parameterized by certain 
deformations of the complex structure of the non compact CY space.
Actually we will elaborate a more general framework.
We consider three dimensional Calabi-Yau geometries
built around a generic Riemann surface $\Sigma$ of any genus which is then the
non trivial 2-cycle we wrap the D-branes around.
The normal bundle is specified by assigning a rank two holomorphic vector
bundle $V$ over $\Sigma$ and the CY condition constrains
the determinant line bundle to equal the canonical line bundle of $\Sigma$,
while $V$ is otherwise generic.

It is in this generic setup that we study the problem of the reduction of the
holomorphic Chern-Simons action functional to the D-brane world volume.
This depends on the $(0,1)$-part of a connection on a $U(N)$ gauge bundle $E$
which we take to be trivial.
Due to the non triviality of the geometry of the normal bundle, in order to specify
the reduction mechanism, we will need to choose a trivialization of the bundle
by a reference non-degenerate bilinear structure $K$ over it.
We will show that choosing the reference connection
to be the (generalized) Chern connection of the bilinear structure $K$
makes the overall result actually independent
both on the reference bilinear form $K$ and on the base representative
of the 2-cycle in the total CY space.
The resulting reduced theory is a generalized
holomorphic $b$--$c$ ($\beta$--$\gamma$) system on $\Sigma$
where the two bosonic fields span a section of $V$.
These are minimally coupled to the reduced gauge connection.

After the above preliminary construction, we consider the effects of varying 
the complex structure
of the total space. Actually we will study constrained variations leaving
the complex structure on $\Sigma$ fixed and preserving the CY condition.
These can be seen to be parametrized by a set of geometric potential 
functions on the
double intersections which are the $\check {\rm Cech}$ cohomology 
representatives of the
complex moduli we are varying. These, in the spirit of Kodaira and Spencer,
can be used to parametrize local singular coordinate changes which specify
the variation of the complex structure.

In order to be able to explicitly deal with the moduli space of the conifold 
complex structures, we then limit
ourselves to the genus zero case, i.e. $\Sigma={\mathbb P}^1$. In this case, the cycle has
a single complex structure and so the above analysis is enough to cover the full moduli
space.
Since the transverse fields are sections of the normal bundle $V$, the singular
coordinate transformation defines the deformation of the reduced theory action
in a well defined way which is parametrized by the geometric potential at the
intersection of the
north and south pole. We calculate this explicitly in the generic case
of CY deformations of the ${\cal O}(n)\oplus{\cal O}(-n-2)$ reference complex structure
on the conifold. The result we find is that the partition function generically 
reduces to an
(n+1)-matrix model whose potential is obtained from the geometric potential
in a specific way.

Let us stress that the above result with $\Sigma={\mathbb P}^1$ and $n=0$ was 
presented in \cite{Dijk1}
where it was suggested that it can be obtained by refining a sketchy 
calculation in \cite{Kachru}. The result presented here 
is a generalization thereof for Riemann surfaces of arbitrary genus and,
for arbitrary
$n$, on ${\mathbb P}^1$.

The same reduction method can be applied also to non compact local CY 
geometries build
around a 4-cycle. In such a case the CY condition specifies the normal line 
bundle to be the canonical line bundle on the base complex manifold. In the 
appendix, we elaborate it in a generic case and find the
reduced holomorphic Chern-Simons theory in the form of an holomorphic BF model
(see \cite{baulieu} for recent discussions on this model).
It was recently suggested in \cite{Diaconescu} that it describes the 
topological open strings for D-branes wrapped around the above four cycles.

\subsection{Reduction in the linear case}

Let us consider non compact six dimensional geometries built around a Riemann 
surface $\Sigma$ as
the total space of a rank $2$ holomorphic vector bundle $V$
with $GL(2,{\mathbb C})$ structure group.

Any atlas $\{ U_{(\alpha)} \}$
on $\Sigma$ extends to an atlas on $CY(\Sigma,V)$ 
by $\hat U_{(\alpha)} = U_{(\alpha)}\times {\mathbb C}^2$.
The complex manifold is defined by the overlapping conditions 
$$
z_{(\alpha)}=f_{(\alpha)(\beta)}\left(z_{(\beta)}\right) \\
$$
\begin{equation}
w^i_{(\alpha)}=M^i_{j(\alpha)(\beta)}\left(z_{(\beta)}\right)
w^j_{(\beta)}
\label{linear}\end{equation}
in any double patch intersection $U_{(\alpha)} \cap U_{(\beta)}$.

Requiring the complex manifold to be of the Calabi-Yau type,
restricts $\det V$ to be equal to the canonical line bundle
$T^{(1,0)}(\Sigma)$ so that 
under this condition the total space of $V$ is equipped with the holomorphic 
$(3,0)$-form $\Omega=dz\wedge dw^1\wedge dw^2$, where $z$ is a local
coordinate system on $\Sigma$ and $w^i$ on the ${\mathbb C}^2$ fibers.
This condition is just $det M_{(\alpha)(\beta)} \times f'_{(\alpha)(\beta)}=1$ 
and it is consistent
with triple intersection conditions.
We denote this manifold $CY(\Sigma,V)$. 

Let us consider the topological open B-model on $CY(\Sigma,V)$, which can be
obtained starting from open string field theory, \cite{Witten}.
Because of the drastic reduction of the degrees of freedom due to the 
huge gauge symmetry present in the topological string, the string field theory 
is the holomorphic Chern-Simons (hCS) theory on $CY(\Sigma,V)$ for a (0,1)-form 
connection on a $U(N)$ bundle $E$, where N is the number of D-branes wrapped
around $\Sigma$. For simplicity, we will restrict to the case in which $E$ 
is trivial. The action of hCS is
\begin{equation}
S(\A)=\frac{1}{g_s}\int_{CY(\Sigma,V)} {\cal L}, 
\quad
{\cal L}=
\Omega\wedge Tr\left(\frac{1}{2} \A\wedge
\bar\partial \A + \frac{1}{3} \A \wedge \A \wedge \A\right)
\label{hCS}\end{equation}
where $\A\in T^{(0,1)}\left(CY(\Sigma,V)\right)$.

 The dynamics of D-branes wrapped around the 2-cycle $\Sigma$
can be described by reducing the open string field theory from the total space 
manifold to the D-brane world-volume.
Since the bundle $V$ is non trivial,
in order to properly define the brane theory, the reduction of the lagrangian
has to be coherently prescribed patch by patch by a trivialization procedure
in such a way that the end product is independent upon the particular 
trivialization we use.

As it is evident, the embedding equations for $\Sigma$ in 
$CY(\Sigma,V)$ are just $w^i=0$. In any local chart, the fibering 
structure defines a local notion of parallel and
transverse directions along which we split 
$\A=\A_{\bar z}d\bar z +\A_{\bar i}d\bar w^i$.
The parallel part $\A_{\bar z}$ glues on double patches intersections as an 
invariant (0,1)-form on $\Sigma$ only when restricted
to the base, while otherwise gets also a linear contribution in
$w$ due to the generic non triviality of $V$
The transverse coefficients $\A_{\bar i}$ glue\footnote{We denote by $V^*$ 
the dual vector bundle, glueing with $(M^{-1})^t$ 
and by $\bar V$ the complex conjugate one.} as
a section of $\bar V^*$. 

Because of this, since the reduction to the base has to be performed 
covariantly, we 
have to expand $\A_{\bar z}=A_{\bar z}-
A_{\bar k}\Gamma_{\bar z \bar j}^{\bar k}\bar w^{\bar j}$
and $\A_{\bar i}=A_{\bar i}$, where $A_{\bar z}d\bar z\in T^{(0,1)}(\Sigma)$,
$A_{\bar i}\in \bar V^*$ and 
$d\bar z\Gamma_{\bar z \bar j}^{\bar k}$ is the $(0,1)$ component of a reference 
connection of $\bar V$.
The reduction process is defined by specifying the subfamily of ${\cal A}$
connections we limit our consideration to. Our prescription is that the 
matrix valued dynamical fields $(A_{\bar z},A_{\bar i})$ that survive the
reduction are those independent of the coordinates along ${\mathbb C}^2$.
A direct calculation from the lagrangian ${\cal L}$ in (\ref{hCS})
for the above reduced configurations gives
\begin{equation}
L=\Omega\wedge Tr
\left(\frac{1}{2}
\left\{
A_{\bar i}D_{\bar z} A_{\bar j}+A_{\bar i}
\Gamma_{\bar z\bar j}^{\bar k}A_{\bar k}
\right\}\right)
d w^{\bar i}\wedge d\bar z\wedge d w^{\bar j}
\label{redlag}\end{equation}
where $D_{\bar z}$ is the covariant derivative w.r.t. the gauge structure.
Notice that the above does not depend on the base representative, that is 
on the values of $w^i$.

Another way of justifying the reduced lagrangian (\ref{redlag}) is the 
following. We start from (\ref{hCS}) and replace the exterior differential 
$\bar\d$ by the covariant differential $\bar\ED= \bar \d + \Gamma$ on 
$CY(\Sigma,V)$ and impose 
that $\ED_{\bar i} {\A_{\bar j}}= 0 = \ED_{\bar i} {\A_{\bar z}}$. The latter
conditions are satisfied as follows. Due to the local product structure of 
$CY(\Sigma,V)$, we can suppose without loss of generality that the only
nontrivial component of the connection relevant to the problem are 
$\Gamma_{\bar z\bar j}^{\bar k}=\Gamma_{\bar j\bar z}^{\bar k}$.
Therefore $\ED_{\bar i} {\A_{\bar j}}= 0$ simply means that ${\A_{\bar j}}$ does
not depend on $w^i$, while $ \ED_{\bar i} {\A_{\bar z}}=0$ can be integrated
and leads precisely the expression for ${\A_{\bar z}}$ given above.

Eq.(\ref{redlag}) represents a six--form. Our purpose is to restrict it to a
two form defined on $\Sigma$.  This has to be consistently prescribed patch 
by patch. That is, we have to couple the above reduction with the contraction 
of the differentials along the fiber directions to obtain a well defined 
$(1,1)$-form on $\Sigma$. To define this operation, let 
us consider a bilinear structure $K$ in $V$, that is
a local section $K\in \Gamma(V\otimes \bar V)$, the components $K^{i \bar j}$
being an invertible complex matrix at any point.

The derivation of the basic $(1,1)$-form is realized patch by patch 
with the help of $K$ as contraction of the 
hCS (3,3)-form Lagrangian by the two bi-vector fields 
$k=\frac{1}{2}
\epsilon_{ij}K^{i \bar l}K^{j \bar k}
\frac{\partial}{\partial \bar w^l}
\frac{\partial}{\partial \bar w^k}$ 
and $\rho=\frac{1}{2}
\epsilon^{ij}\frac{\partial}{\partial w^i}
\frac{\partial}{\partial w^j}$.
Notice that $k\in det V$ and $\rho\in det V^*=(det V)^{-1}$
so that the combined application of the two is a globally well--defined
operation.
Calculating then the pullback Lagrangian, we obtain
\begin{equation}
{\cal L}_{red}= i_{\rho\wedge k} L=
\um dz d{\bar z} (det K) 
\epsilon^{\bar i\bar j}
Tr\left[
A_{\bar i}D_{\bar z}A_{\bar j} + 
A_{\bar i}\Gamma_{\bar z\bar j}^{\bar k}A_{\bar k}
\right]
\label{pb'}\end{equation}

Our last step relates the reference connection and the reference bilinear 
structure in order to obtain a result which is independent upon the 
trivialization we used. Define the field components $\varphi^i=i_{V^i}A\in V$, 
where $V^i=K^{i \bar j}\frac{\partial}{\partial\bar w^j}$
and plug it in (\ref{pb'}). One gets
\begin{equation}
{\cal L}_{red}= 
\um dz d{\bar z}  
Tr\left[\epsilon_{ij}
\varphi^iD_{\bar z}\varphi^j + (det K)
\varphi^m\varphi^n \epsilon^{\bar i\bar j}
\left(
K_{m \bar i}\partial_{\bar z}K_{n \bar j}
+K_{m \bar i}K_{n \bar K}
\Gamma_{\bar z\bar j}^{\bar k}
\right)\right]
\label{pb''}\end{equation}
where $K_{\bar i j}$ are the components of the inverse bilinear structure, that is
$K_{\bar i j} K^{j \bar l}=\delta_{\bar i}^{\bar l}$.
In order to have a result which is independent on the trivialization, just
set the reference connection to be the generalized Chern connection of 
the bilinear structure $K$, that is
$\Gamma_{\bar z\bar j}^{\bar k}= K_{\bar j l}\partial_{\bar z}K^{l\bar k}$.
Therefore, choosing our reference trivialization $(\Gamma,K)$ data 
to satisfy this natural condition, we get 
\begin{equation}
{\cal L}_{red}= \um
dz d{\bar z}Tr\left[
\epsilon_{ij}\varphi^iD_{\bar z}\varphi^j 
\right]
\label{pb}\end{equation}
which is a well defined $(1,1)$-form on $\Sigma$. Hence the
action for the reduced theory is given by
$$
S_{red}=\frac{1}{g_s}\int_\Sigma {\cal L}_{red}=
\frac{1}{g_s}\um\int_\Sigma dz d{\bar z}Tr\left[
\epsilon_{ij}\varphi^iD_{\bar z}\varphi^j 
\right].
$$

\subsection{Deformations of the complex structure}

Let us now discuss certain variations of the complex structures of the 
manifold $CY(\Sigma,V)$ following the approach of Kodaira.

A general complex structure variation of the vector bundle structure 
(\ref{linear}) is given by the deformed patching conditions
$$
\xi_{(\alpha)}=f_{(\alpha)(\beta)}\left(\xi_{(\beta)}\right) + 
\delta_{(\alpha)(\beta)}\left(\xi_{(\beta)},\omega_{(\beta)}\right)
$$
\begin{equation}
\omega^i_{(\alpha)}=M^i_{j(\alpha)(\beta)}\left(\xi_{(\beta)}\right)
\left[
\omega^j_{(\beta)} + \Psi^j_{(\alpha)(\beta)}\left(\xi_{(\beta)},\omega_{(\beta)}\right)
\right]
\end{equation}
where $\alpha$ and $\beta$ label the two local charts, and 
$\delta$ and $\Psi^i$ are analytic functions on double patch intersections.
The variation is trivial if it can be re-absorbed via an analytic change of 
coordinates. Notice that in the general case, the deformation functions are 
constrained by the chain rules of multiple patch intersection.

In the following, we will consider variations leaving invariant the complex 
structure on $\Sigma$. It is obvious that this coincides with the general 
case if the moduli space of complex structures of $\Sigma$ is a point. 
Then, from now on, we will restrict to variations of the form
$$
z_{(\alpha)}=f_{(\alpha)(\beta)}(z_{(\beta)}) 
$$
\begin{equation}
\omega^i_{(\alpha)}=M^i_{j(\alpha)(\beta)}\left(z_{(\beta)}\right)
\left[
\omega^j_{(\beta)} + \Psi^j_{(\alpha)(\beta)}\left(z_{(\beta)},
\omega_{(\beta)}\right)
\right]
\end{equation}
Notice that the deformed complex structure preserves the CY condition if
in any $U_{(\alpha)} \cap U_{(\beta)}$ we have
$det\left(1+\partial\Psi\right)=1$, where
$(1+\partial\Psi)^i_j=\delta^i_j+ \partial_j\Psi^i$.

The solution of the above CY condition can be easily given in terms of a
set of potential functions (one for each double patch intersection modulo
triple intersections identities) which generates the deformation, as 
$$
\epsilon_{ij}
w^i_{(\alpha)(\beta)}d w^j_{(\alpha)(\beta)}
=
\epsilon_{ij}
\omega^i_{(\alpha)}d \omega^j_{(\alpha)} - d X^{(\alpha)(\beta)},
$$ 
where we defined
$w^i_{(\alpha)(\beta)}=\omega^i_{(\alpha)}+
\Psi^i_{(\alpha)(\beta)}(z^{(\beta)},\omega_{(\beta)})$.

For later application, let us specify the previous general construction for
$\Sigma={\mathbb P}^1$. The patching on the sphere allows a 
drastic simplification of the above formulas.
In this case the moduli space of complex structure of the base Riemann surface
is point-like, so keeping it fixed is not a constraint.
The sphere can be described by the standard charts $U_{S/N}$ around the 
north and south poles and the single intersection 
$U_S\cap U_N={\mathbb C}^\times$ is the cylinder ${\mathbb C}\setminus\{0\}$.

As it is well known, by Grothendieck's theorem, any holomorphic vector bundle 
on ${\mathbb P}^1$ can be presented as a direct sum of line bundles. In our case
therefore $V={\cal O}(-n_1)\oplus {\cal O}(-n_2)$,
where we denote by ${\cal O}(-n)$ the line bundle defined by the glueing 
rules
$$
z_N=-z_S^{-1}
\quad{\rm and}\quad
w_N=z_S^n w_S.
$$
The CY condition for the total space $CY(\Sigma,V)$ is therefore
$n_1+n_2=2$.

The generic variation is 
\begin{equation}
\omega_N^i=z_S^{n_i}\left[\omega_S^i+\Psi^i\left(z_S,\omega_S\right)\right]
\label{var}\end{equation}
where $\omega_S=(\omega^1_S,\omega^2_S)$.
In Eq.(\ref{var}), since there is one single double intersection and no 
triple ones, the functions $\Psi^i$ are just constrained to be analytic
on ${\mathbb C}^\times \times {\mathbb C}^2$, that is are allowed to have poles of finite
orders at $0$ and $\infty$ and have to be analytic in $\omega_S$. 
The relevant terms in $\Psi^i$, i.e. the ones representing true variations
of the complex structure of the initial space, are the ones which cannot be
re-absorbed by analytic reparametrizations of $\omega_S$ and $\omega_N$.
Moreover, the Calabi-Yau condition in the new complex structure is solved by a single
potential function $X=X(z_S,\omega_S)$ such that
\begin{equation}
\epsilon_{ij}w^id w^j
=
\epsilon_{ij}\omega^id \omega^j - d X,
\label{varvar}\end{equation}
where, as before,
$w^i=\omega^i+\Psi^i(z_S,\omega)$.

Let us note now that we can, as it is usually done in Kodaira-Spencer theory,
relate the deformed and the original complex structures by a singular 
change of coordinates. For the case at hand, it is enough to do it along 
the fibers above the south pole patch, namely
\begin{equation}
w^i_N=\omega^i_N,\quad {\rm and} \quad
w^i_S=\omega^i_S+\Psi^i\left(z_S,\omega_S\right)
\label{singular}\end{equation}
In the singular coordinates $(z,w^i)$ the patching rule is the original
linear one. Therefore, Eq. (\ref{singular}) defines naturally 
the transformation rule for generic sections in the deformed complex structure
from the singular to the non singular coordinate system.

\subsection{Reduction over $\Sigma={\mathbb P}^1$ in the deformed case}

Let us now perform the reduction to the brane of the open string field theory
action on a Calabi-Yau deformation of
$CY(\Sigma,V)$ with $\Sigma={\mathbb P}^1$ and $V={\cal O}(-n_1)\oplus{\cal O}(-n_2)$
with $n_1+n_2=2$. 
Actually, from the perspective we adopted so far it turns out that performing it
is not crucially different from the linear case.
That's because we can proceed by performing the reduction of the hCS theory
in
the singular coordinates
(\ref{singular}) 
following the prescription proper to
the linear undeformed case and then implement the variation of the complex structure
by passing to the non singular variables by the proper field redefinition.

\vspace{.3cm}

Let us start for simplicity with the reduction in {\it the Abelian $U(1)$ case}.
In this case the cubic term in the hCS Lagrangian is absent
and the reduction is almost straightforward. In the singular coordinates 
we obtain that
\begin{equation}
{\cal L}_{red}=\um\epsilon_{ij}\varphi^i\partial_{\bar z}\varphi^j dz d{\bar z}
\label{pbsing}\end{equation}
in both the north and south charts. 
The coordinate change for the fields $\varphi^i$ in terms of the ones
corresponding to the deformed complex structure is induced by
(\ref{singular}). 
Let us recall that the functions $\Psi^i$ defining the deformation
are built out from the potential $X$ as in eq. (\ref{varvar}).
This expresses exactly our Lagrangian terms (patch by patch) 
\begin{equation}
\epsilon_{ij}\varphi^i\partial_{\bar z} \varphi^j
=
\epsilon_{ij}\phi^i \partial_{\bar z} \phi^j
- \partial_{\bar z} X
\label{fields}\end{equation}
where $X_N=0$ and $X_S$ is an arbitrary analytic function of the $\phi$'s
in ${\mathbb C}^2$ and of $z$ in ${\mathbb C}^\times$ ($\varphi^i$ are akin to
the coordinate singular coordinate $w^i$ of the previous subsection, 
while $\phi$ stem from $\omega^i$). 

The above potential term $X$ gives the deformation of the action due to the
deformation of the complex structure. Specifically, we have that 
\begin{equation}
S_{red}=\frac{1}{g_s}
\left[
\int_{U_S} \chi_S  ({\cal L}_{red})_S +
\int_{U_N} \chi_N  ({\cal L}_{red})_N
\right]
\label{abred}\end{equation}
where we explicitly indicated the resolution of the unity on the sphere 
$1=\chi_S+\chi_N$. For simplicity we choose the $\chi$'s to be simply step 
functions on the two hemispheres.
Substituting (\ref{pbsing}) and (\ref{fields}) in (\ref{abred}) we then obtain
\begin{equation}
S_{red}=\frac{1}{2\,g_s}
\left[\int_{P^1} \epsilon_{ij}\phi^i \partial_{\bar z} \phi^j dz d\bar z
- \int_{D} \partial_{\bar z} X(z,\phi) dz d\bar z\right]
\end{equation}
where $D$ is the unit disk (south hemisphere).
The disk integral can be reduced by the Stokes theorem, leaving finally
\begin{equation}
S_{red}=\frac{1}{2\,g_s}
\left[\int_{P^1} \epsilon_{ij}\phi^i \partial_{\bar z} \phi^j dz d\bar z
+ \oint X(z,\phi) dz\right]
\end{equation}
where $\oint$ is a contour integral along the equator (we understand the factor
$1/2pi i$).

Therefore, we see that the reduced theory gives a $b$--$c$ ($\beta-\gamma$) system 
on the two hemispheres with a junction interaction along the equator
and the identifications (\ref{var}) on the fields.

\vspace{.3cm}

{\it The non--Abelian case} is a bit more complicated than the Abelian one 
because of the tensoring with the (trivial) gauge bundle. 
This promotes the vector bundle sections
to matrices and therefore unambiguously defining  
the potential function $X$ in the general case is not immediate.
In the following we show where the difficulty arises and what 
further constraint to the deformation of the complex structure is needed
in order to suitably deal with the non--Abelian case.

To see this let us perform the reduction
on ${\mathbb P}^1$ of the non Abelian hCS (\ref{hCS}), as we did in the Abelian
case. Let us work in the singular coordinates and obtain again the pullback
Lagrangian we got in the linear case. Now, in order to pass to the non singular 
coordinates we have to promote to a matrix equation the change of variables
(\ref{fields}). This can be done by specifying a prescription for matrix 
ordering. Suppose we choose a specific ordering and denote it by $\P$.
Then our change of variable is 
\begin{equation}
Tr[\epsilon_{ij}\varphi^i\partial_{\bar z} \varphi^j]
=
Tr[\epsilon_{ij}\phi^i \partial_{\bar z} \phi^j]
- \partial_{\bar z} Tr X^\P
\label{matfields}\end{equation}
while the cubic term gives
\begin{equation}
Tr \left[A_{\bar z} \epsilon_{ij} (\phi^i +\Psi^{i\P})(\phi^j
+\Psi^{j\P})\right]
\label{cubicte}\end{equation}
It appears immediately that our result is complicated and seems to depend quite
non--trivially on the matrix ordering prescription. 
Otherwise, it is well defined.

The easiest way not to have to do with matrix ordering prescription is to have
to do with only one matrix. Henceforth we restrict to the case in which 
$X(z,\omega)$ does not depend, say, on $\omega_2$ and we proceed further.

In this case the deformation formulas simplify considerably.
Eq.(\ref{varvar}) is solved by $\Psi^1=0$ and $\Psi^2$ is determined by the 
potential by
\begin{equation}
\partial_{\omega^1}\left(\frac{\Psi^2}{\omega^1}\right)=
-\frac{\partial_{\omega^1} X}{(\omega^1)^2}
\label{geomat}\end{equation}
This condition can be written also as
$$2\Psi^2=\partial_{\omega^1}\left[\omega^1\Psi^2+X\right].$$

As far as the reduction is concerned, eq.(\ref{matfields}) is unchanged,  
since we do not need any prescription $\P$; while eq.(\ref{cubicte}) 
simplify to
\begin{equation}
Tr \left[A_{\bar z} \epsilon_{ij} (\phi^i +\Psi^{i})(\phi^j +\Psi^{j})\right]
=
Tr \left(A_{\bar z}\left[\phi^1,\phi^2\right]\right)
\label{cubicte'}
\end{equation}
(where we used $[\phi^1,\Psi^2]=0$)
which, as in the linear case, is the contribution needed to complete the 
covariant derivative. 
The last operation to obtain our final result is an integration by part in the 
derivative term. As
$$\epsilon_{ij}\phi^i\partial_{\bar z}\phi^j=-2\phi^2\partial_{\bar z}\phi^1 
+\partial_{\bar z}\left(\phi^1\phi^2\right)$$ in both the north and south 
charts, from the last term we get an additional contribution to the equator 
contour integral, that is
$\frac{1}{g_s}\um\oint Tr\phi^1\Psi^2$. Adding it to the previously found 
term we get $\frac{1}{g_s}\um\oint Tr(X+\phi^1\Psi^2)$.
This, by (\ref{geomat}) can be written just as $\frac{1}{g_s}\oint Tr B$, 
where $\partial_\omega^1 B=\Psi^2$.

Therefore, summarizing, we find that in the non Abelian case on the Riemann 
sphere  we are able to treat the deformations of the type
\begin{equation}
\omega_N^1=z_S^{-n}\omega_S^1 ,
\quad{\rm and}\quad
\omega_N^2=z_S^{2+n}\left[\omega_S^2+
\partial_{\omega^1}B\left(z_S,\omega_S^1\right)\right]
\label{varfin}\end{equation}
which corresponds to the choice $n_1=-n$. This geometry has been introduced
(in the matrix planar limit) by \cite{Ferrari}.
These geometries are CY for any potential $B$ analytic in 
${\mathbb C}^\times\times{\mathbb C}$. The relevant reduced theory action 
is given by
\begin{equation}
S_{red}=\frac{1}{g_s}
\left[\int_{P^1} -\Tr(\phi^2 D_{\bar z} \phi^1) dz d\bar z
+ \oint \Tr B(z,\phi^1) dz\right]
\end{equation}

\subsection{Reduction to matrix models}

For completeness, let us generalize to our case the argument of \cite{Aganagic}
to show the reduction of the previous action to the matrix models. 
In calculating the partition function of the open strings 
attached to the D-branes\footnote{Notice that the field redefinition $\varphi\to\phi$
has unit Jacobian because of the CY condition.}
we can easily integrate out the gauge connection $A_{\bar z}$ which implies 
the constraint $[\phi^1,\phi^2]=0$ and then $\phi^2$ which implies the constrain 
$\partial_{\bar z}\phi^1=0$.  As a result\footnote{See   
 \cite{Marino}, for more details about how the ghost contribution in the 
 maximal Abelian gauge  
compensates the $det^{-1} ad_{\phi^1}$ term.} we get that the partition function 
\begin{equation}
Z_{red}=
\int D[A_{\bar z},\phi^1,\phi^2]e^{-S_{red}}
\propto
\int D[\phi^1]\delta(\partial_{\bar z}\phi^1)e^{-\frac{1}{g_s}\oint B}
\label{pf1}\end{equation} 

The condition such that the equation $\partial_{\bar z}\phi^1=0$ admits 
solutions is 
$n\geq0$ in (\ref{varfin}). In such a case we have $n+1$ independent solutions
which in the south patch are the linear span of $\{z_S^i\}_{i=0\dots n}$.

Therefore, expanding $\phi^1_S=\sum_{i=0}^n X_i z_S^i$ in (\ref{pf1}),
we are left with the multi--matrix model partition function
\begin{equation}
Z_{red}= \int\prod_i dX_i e^{-W(X_0,\dots X_n)}
\label{mmpf}\end{equation}
where the potential is given by
\begin{equation}
W(X_0,\dots X_n)=\oint dz B\left(z, \sum_{i=0}^n X_i z^i\right)
\label{mmpot}\end{equation}
This coincides with the one introduced by \cite{Ferrari} in the matrix 
planar limit.

The original Dijkgraaf-Vafa case is reproduced for $n=0$. Then, the only non 
trivial complex structure deformation in (\ref{varfin}) is with 
$B=\frac{1}{z}W(\omega_1)$ (since any other dependence in $z$ can be 
re-absorbed by analytic reparametrizations)
and hence we get the one matrix model with potential $W$.

The above formula can be also inferred by just generalizing 
another CFT argument by Dijkgraaf-Vafa to the geometry (\ref{varfin}).
To this end let us consider again the two dimensional theory defined by the 
action
\begin{eqnarray}
S = \frac{1}{g_s} \int_{{\mathbb P}^1} \Tr \left( \phi_2 D_{\bar z} \phi_1 
\right)
\end{eqnarray}
where $D_{\bar z} = \partial_{\bar{z}}+[A_{\bar{z}}, \cdot ]$. This is a 
gauged chiral conformal field theory: 
a gauged $b$--$c$ ($\beta$--$\gamma$) system in which $\phi_1$ and $\phi_2$ are 
conformal fields of dimensions $-n/2$ and $1+n/2$ respectively. 
For any $n$, the fields $\phi_1$ and $\phi_2$ are canonically conjugated and 
on the plane they satisfy
the usual OPE \begin{eqnarray}
   \phi_1 (z) \phi_2 (w) \sim \frac{g_s}{z-w}
\end{eqnarray}
In Hamiltonian formalism, that is in the radial quantization of the CFT, the 
partition function is given as
\begin{eqnarray}
Z = \left< \mathrm{out} | \mathrm{in} \right>.
\end{eqnarray}
The deformed transformation
\begin{eqnarray}
%\phi_1' &=& z^{-n} \phi_1 \nonumber \\
\phi_2' &=& z^{n+2} \left( \phi_2 + \partial_{\phi_1} B(z,\phi_1) \right)
\end{eqnarray}
is given on the cylinder $z=e^{w}$ as
\begin{eqnarray}
\left( \phi_2'\right)_{cyl} &=& \left(\phi_2\right)_{cyl}
                               + \frac{\partial B(z,\phi_1) }{\partial 
{\left(\phi_1\right)_{cyl}}}
\end{eqnarray}
and is implemented by the operator
\begin{eqnarray}
U = \exp \left(\Tr \oint \frac{\de z}{2 \ii \pi} B(z,\phi_1)  \right)
\end{eqnarray}
Therefore the new partition function is
\begin{eqnarray}
Z = \left< \mathrm{out} |U| \mathrm{in} \right>.
\end{eqnarray}
which is our result.

We remark that this is an {\it a posteriori} argument, it is a consistency check
but does not explain the dynamical origin of the matrix model from the string 
theory describing the brane dynamics.

\section{Engineering matrix models}

Once the link between D-brane configurations and multi--matrix models is 
established, the next natural question to ask is what kind of matrix 
models we get in this way. In this section we single out what is the most 
general type of multi--matrix model
we can engineer by deforming D-branes on 2--cycles in the above way and 
we produce some examples. 

The geometric potential $B(z,\omega)$ is a general holomorphic function on 
$\mathbb{C}^* \times \mathbb{C}$ 
but the terms actually contributing to a change in the complex structure 
and giving a non zero matrix potential
are of the form
\begin{eqnarray}\label{eq:geometricpotential}
B(z, \omega) = \sum_{d=1}^{\infty}  \sum_{k=0}^{d \cdot n} t^{(k)}_{d} 
z^{-k-1} \omega^d
\end{eqnarray}
where $t^{(k)}_{d}$ are the \emph{times} of the potential and $\omega$ is to 
be identified with the coordinate $\omega_1$ of the previous section.
It can be easily proven that other terms in the expansion can be re-absorbed 
by an analytic change of coordinates in the geometry.
Consistently with the geometric theory of deformations \cite{katz}, they 
do not contribute to the matrix potential.

The \emph{degree} of the potential $B$ is the maximum $d$ such that 
$ t^{(k)}_{d}$ is non-zero for some $k$ in (\ref{eq:geometricpotential}) 
and corresponds to the degree of the matrix potential, obtained as
\begin{eqnarray}\label{eq:matrixpotential}
% \phi^* = \sum_{j=0}^{n} x_i z^i \\
W(X_0, \dots, X_n) = {\oint} \frac{\de z}{2 \ii \pi} 
B \left( z, \sum_{j=0}^{n} X_i z^i \right).
\end{eqnarray} 
Since this operation is linear, 
from (\ref{eq:geometricpotential}) and (\ref{eq:matrixpotential}) one gets 
a matrix potential of the form
\begin{eqnarray}\label{eq:matrixpotential2}
% \phi^* = \sum_{j=0}^{n} x_i z^i \\
W(X_0, \dots, X_n) =   \sum_{d=0}^{\infty}  \sum_{k=0}^{d \cdot n} 
t^{(k)}_{d} W^{(k)}_{d}(X_0, \dots, X_n)
\end{eqnarray} 
where each term
\begin{eqnarray}
 W^{(k)}_{d}(X_0, \dots, X_n) = \sum_{i_1, \dots, i_d=0 \atop i_1+ \dots + 
 i_d=k }^n X_{i_1} \dots X_{i_d}
\end{eqnarray}
corresponds to $ B^{(k)}_{d} (z, \omega) = z^{-k-1} \omega^d$ for 
$0 \leq d \leq + \infty$ and $0 \leq k \leq{d \cdot n}$. Note that these are 
directly obtained in completely symmetric ordered form with respect to 
the indices $i_1, \dots i_d$ labeling the different matrix variables. 
In the following we will sometimes write simply the polynomial $W$ for 
c-number variables $W(x_0, \dots, x_n)$, understanding 
the total symmetrization when matrices are plugged in.

As it was already anticipated in the previous section, the {\it one matrix 
models} corresponds directly to the Dijkgraaf-Vafa
case with $n=0$ and therefore $B=\frac{1}{z}W(\omega)$. 

{\it Two matrix models} are obtained by considering the case $n=1$.
Some of them have been derived in \cite{Ferrari}.
In this case, it is possible to engineer a general function for two commuting variables. In fact
\begin{eqnarray}
B(z, \omega) = z^{-k-1} \omega^{k+j} & \rightarrow & W(x)= 
{k+j \choose k} x_0^k x_1^j .% \qquad 0 \leq k \leq n
\end{eqnarray} 
and the matrix potential reads
\begin{eqnarray}
W(x_0,x_1)=
\sum_{d=1}^\infty
\sum_{k=0}^d
t_d^{(k)}{d \choose k}
x_1^k x_0^{d-k}
\end{eqnarray} 
which is, upon varying the possible couplings, a generic analytic potential 
in the two variables $x_0$ and $x_1$.
The only constraint is the matrix ordering which is always the symmetric one.
In particular, it is easy to engineer a two matrix model with bilinear 
coupling.  This is achieved by choosing, for $n=1$,
the geometric potential to be
\begin{eqnarray}
B(z, \omega) = \frac{1}{z} \left[V(\omega)+U\left(\frac{\omega}{z}\right)\right]
+\frac{c}{2z^2}\omega^2\label{2mpot}
\end{eqnarray} 
which generates the matrix potential
\begin{eqnarray}
W(x_0,x_1) = V(x_0)+U(x_1)+cx_0x_1 \label{2mpot'}
\end{eqnarray} 

In general, the {\it multi--matrix models} one can engineer are not   
of arbitrary form. Actually, on top of the fact that we can generate only 
matrix potentials with symmetric ordering, there are also constraints between 
possibly different couplings. This can be inferred from the fact that a 
polynomial function in $n+1$ variables  of total maximal degree $d$ is 
specified by many more coefficients than the ones we have at our disposal. 
(As an example, if $n=3$ and $d=3$ we would need $10$ coefficients, 
while we have only $4$ at our disposal.)

To end this section, we would like to remark that some deformations can connect 
cases with different values of $n$. The geometric equivalence of seemingly 
different complex structures becomes in fact explicit at the matrix model level.
As an example, let us consider
the case $n=2$ and a geometric potential of the form
\begin{eqnarray}
B(z, \omega) = -\frac{1}{2} z^{-4} \omega^2 + z^{-3} F(\omega).\0
\end{eqnarray} 
Out of this, one obtains
\begin{eqnarray}
W(x_0, x_1, x_2) = \left( F'(x_0) - x_1\right) x_2 + 
\frac{1}{2}\left( F''(x_0)x_1^2 \right)\0
\end{eqnarray} 
After integration of $x_2$, which appears linearly, this theory is equivalent 
to a one-matrix model with potential
\begin{eqnarray}
V(x_0) = \frac{1}{2}F''{F'}^2(x_0),\0
\end{eqnarray} 
which is equivalent to $n=0$ and $B(z,\omega)=\frac{1}{z}V(\omega)$.
As a matter of fact, the geometry with $n=2$ and $B=-\frac{1}{2} z^{-4} \omega^2 $ is 
equal, upon diagonalization,
to the geometry $n=0$ and $B=0$.

\section{General properties of two--matrix models.}

\setcounter{equation}{0}
\setcounter{footnote}{0}

In the second part of the paper we concentrate on a subclass of matrix models:
the two--matrix models with bilinear coupling between the two matrices but 
arbitrary self--coupling of each matrix. Our purpose is to find {\it exact
quantum} solutions. For this reason we will solve them with the method of 
orthogonal polynomials. This method allows one to explicitly perform the path 
integration, so that one is left with quantum equations. The two basic 
ingredients are the quantum equations of motion and the integrable linear 
systems. The latter in particular
uncover the integrable nature of two--matrix models, which stems from the Toda 
lattice hierarchy \cite{UT} underlying all of them. 
Our approach for solving two--matrix 
models consists in solving the quantum equations of motion and, then, using the
recursiveness intrinsic to integrability (the flow equations), finding explicit 
expressions for the correlators. An alternative method 
is based on the W constraints on the functional integral.
We do not use it here, but one can find definitions, applications and 
comparisons with the other methods in \cite{BX1,BX2,BX3,BCX}.

For general reviews on matrix models applied to string theory, see
\cite{Kleb,GM,DFGZ}. For general application of matrix models, see 
\cite{Kostov1,kazakov}. Early literature on two--matrix models is contained in
refs.\cite{Douglas} through \cite{Stau}. The method used in the present paper,
although implicit in the early literature, is, quite incomprehensibly, 
seldom utilized. Different methods (from the saddle point to 
loop equations) are often preferred, see  refs.\cite{Akemann} through  
\cite{Alex2}.

\subsection{Review of old formulas}

The model of two Hermitian $N \times N$ matrices $M_1$ and $M_2$ with bilinear
coupling, see (\ref{2mpot},\ref{2mpot'}), is introduced in terms of the 
partition function 
\a
Z_N(t,c)=\int dM_1dM_2 e^{tr W},\quad\quad
W=V_1 + V_2 + c M_1 M_2\label{Zo}
\b
with potentials
\a
V_{\al}=\sum_{r=1}^{p_\al} \bar t_{\al,r}M_{\al}^r\,\qquad \al=1,2.\label{V}
\b
where $p_\al$ are finite numbers. These potentials define the model.
We denote by ${\cal M}_{p_1,p_2}$ the corresponding two--matrix model. 

We are interested in computing correlation functions (CF's) of the operators
\a
\tau_k=tr M_1^k,\qquad \sigma_k=tr M_2^k ,\qquad
\forall k,\0
\b
For this reason we complete the above model by replacing
(\ref{V}) with the more general potentials
\a
V_\al = \sum_{r=1}^\infty  t_{\al,r} M_\al^r, \qquad \al =1,2\label{Vgen}
\b
where $t_{\al , r} \equiv \bar t_{\al,r}$ for $r\leq p_\al$. The CF's are
defined by
\a
< \tau_{r_1}\ldots \tau_{r_n}\sigma_{s_1}\ldots \sigma_{s_m}> =
\frac {\d^{n+m} }{\d t_{1,r_1}\ldots \d t_{1,r_n}\d t_{2,s_1}\ldots 
\d t_{2,s_m}} \ln Z_N(t,g)\label{CF}
\b
where, in the RHS, all the $t_{\al,r}$ are set equal to $\bar t_{\al,r}$
for $r\leq p_\al$ and the remaining are set to zero. The unusual + sign at 
the exponent of the integrand in (\ref{Zo}) is because we
want to use a uniform notation for physical couplings  $\bar t_{\al,r}$
and sources $t_{\al,r}$ (for the convergence of the integrals, see below).   
From now on we will not distinguish between $t_{\al,r}$ and $\bar t_{\al,r}$
and use throughout only $t_{\al,r}$. We hope the context will always make clear
what we are referring to. 

We recall that the ordinary procedure to calculate the partition function 
consists of three steps \cite{BIZ},\cite{IZ2},\cite{M}:
$(i)$ one integrates out the angular part so that only the
integrations over the eigenvalues are left;
$(ii)$ one introduces the orthogonal monic polynomials
\a
\xi_n(\lambda_1)=\lambda_1^n+\hbox{lower powers},\qquad\qquad
\eta_n(\lambda_2)=\lambda_2^n+\hbox{lower powers}\0
\b
which satisfy the orthogonality relations
\a
\int  d\lambda_1d\lambda_2\xi_n(\lambda_1)
e^{V_1(\lm_1)+V_2(\lm_2)+c\lm_1\lm_2}
\eta_m(\lambda_2)=h_n(t,c)\delta_{nm}\label{orth1}
\b
$(iii)$, using the orthogonality relation (\ref{orth1}) and the properties
of the Vandermonde determinants, one can easily
calculate the partition function
\a
Z_N(t,c)={\rm const}~N!\prod_{i=0}^{N-1}h_i\label{parti1}
\b
whereby we see that knowing the partition function means knowing
the coefficients $h_n(t,c)$.

The crucial point is that the information 
concerning the latter can be encoded in 1) a suitable linear system 
subject to certain 2) equations of motion (coupling conditions),
together with 3) relations that allows us to reconstruct $Z_N$. 

Let us introduce some convenient notations. We will meet infinite matrices 
$M_{ij}$ with $0\leq i,j <\infty$. For any such matrix $M$, we define
\a
\cm=H^{-1} MH,\qquad H_{ij}=h_i\delta_{ij},\qquad \widetilde M_{ij} = M_{ji},
\qquad M_l(j)\equiv M_{j,j-l}.\0
\b
We represent such matrices in the lower right quadrant of the $(i,j)$ plane.
They all have a band structure, with nonzero elements belonging
to a band of lines parallel to the main descending diagonal. 
We will write $M\in [a,b]$, if all its non--zero lines are between
the $a$--th and the $b$--th ones, setting $a=0$ for the main diagonal.
Moreover $M_+$ will denote the upper triangular
part of $M$ (including the main diagonal), while $M_-=M-M_+$. We will write
\a
{\rm Tr} (M)= \sum_{i=0}^{N-1} M_{ii}\0
\b
 
Next we pass from the basis of orthogonal polynomials
to the basis of orthogonal functions
\a
\Psi_n(\lambda_1)=e^{V_1(\lambda_1)}\xi_n(\lambda_1),
\qquad
\Phi_n(\lambda_2)=e^{V_2(\lambda_2)}\eta_n(\lambda_2).\0
\b
The orthogonality relation (\ref{orth1}) becomes
\a
\int d\lm_1 d\lm_2\Psi_n(\lambda_1)e^{c\lm_1\lm_2}
\Phi_m(\lambda_2)=\delta_{nm}h_n(t,c).\label{orth2}
\b
We will denote by $\Psi$ the semi--infinite column vector
$(\Psi_0,\Psi_1,\Psi_2,\ldots,)^t$ and by $\Phi$ the vector  
$(\Phi_0,\Phi_1,\Phi_2,\ldots,)^t$.
Then we introduce the following $Q$--type matrices
\a
\int d\lm_1 d\lm_2\Psi_n(\lambda_1)
\lm_{\al}e^{c\lm_1\lm_2}
\Phi_m(\lambda_2)\equiv Q_{nm}(\al)h_m=\q_{mn}(\al)h_n,\quad
\al=1,2.\label{Qalpha}
\b

Beside these $Q$ matrices, we will need two $P$--type matrices, defined by
\a
&&\int d\lm_1 d\lm_2\Bigl(\ddlm 1 \Psi_n(\lambda_1)\Bigl)
e^{c\lm_1\lm_2}\Phi_m(\lambda_2)\equiv P_{nm}(1)h_m\label{P(1)}\\
&&\int  d\lambda_1d\lambda_2\Psi_n(\lambda_1)e^{c\lm_1\lm_2}
\Bigl(\ddlm 2 \Phi_m(\lambda_2)\Bigl)\equiv P_{mn}(2)h_n\label{P(2)}
\b
For later use we also introduce
\a
&&\int d\lm_1 d\lm_2\Bigl(\ddlm 1 \xi_n(\lambda_1)\Bigl)
e^{V_1(\lm_1)+V_2(\lm_2)+c\lm_1\lm_2}\eta_m(\lambda_2)\equiv 
P^\circ_{nm}(1)h_m\label{P(1)'}\\
&&\int  d\lambda_1d\lambda_2\xi_n(\lambda_1)e^{V_1(\lm_1)+
V_2(\lm_2)+c\lm_1\lm_2}
\Bigl(\ddlm 2 \eta_m(\lambda_2)\Bigl)\equiv P^\circ_{mn}(2)h_n\label{P(2)'}
\b

Let us come now to the three elements announced above.

1) {{\it Quantum equations of motion}}. 
\noindent The two matrices (\ref{Qalpha}) we
introduced above are not independent. More precisely both
$Q(\alpha)$'s can be expressed in terms of only one of them and 
one matrix $P$.
Expressing the trivial fact that the integral of the total derivative of the
integrand in eq.(\ref{orth2}) with respect to $\lm_1$ and $\lm_2$
vanishes, we can easily derive the constraints or {\it coupling conditions, or
quantum equations of motion}
\a
P^\circ(1)+ V_1'+c Q(2)=0,\qquad\quad
c Q(1)+V_2'+\widetilde{\cal P}^\circ(2)=0,\label{coupling}
\b
These may be considered the quantum analog of the classical equations of motion.
The difference with the classical equations of motion of the original matrix
model is that, instead of the $N \times N$ matrices $M_1$ and $M_2$, here we 
have infinite $Q(1)$ and $Q(2)$ matrices together with the quantum deformation 
terms given by $P^\circ(1)$ and ${\cal P}^\circ(2)$, respectively.
From the coupling conditions it follows at once that
\a
Q(\al)\in[-m_{\al}, n_{\al}],\qquad \al=1,2\0
\b
where
\a
m_1=p_2-1, \qquad\quad m_2=1 \qquad\quad\quad
n_1=1, \qquad\quad n_2=p_1-1\0
\b
where $p_\al$, $\al =1,2$ is the highest order of the interacting part of
the potential $V_\al$ (see (\ref{V})).

2) ${{\it The~ associated~~ linear~~ systems}}$. 
\noindent The derivation of the  linear systems associated to our matrix model
is very simple.  We take the derivatives of eqs.(\ref{orth2})
with respect to the time parameters $t_{\al,r}$, and use
eqs.(\ref{Qalpha}).  We get in this way the time evolution of $\Psi$
and $\Phi$, which can be represented in two different ways:

\vskip0.2cm
\noindent
${Discrete~  Linear~ System~~I}$:
\a
\left\{\ba{ll}
Q(1)\Psi(\lambda_1)=\lambda_1\Psi(\lambda_1),& \\\noal
{\partial\over{\partial t_{1,k}}}\Psi(\lambda_1)=Q^k(1)_+
\Psi(\lambda_1),\\\noal
{\partial\over{\partial t_{2,k}}}\Psi(\lambda_1)=-Q^k(2)_-
\Psi(\lambda_1),\\\noal
{\partial\over{\partial\lm}}\Psi(\lambda_1)=P(1)\Psi(\lm_1).&
\ea\right.\label{DLS1}
\b
The corresponding consistency conditions are
\ai
&&[Q(1), ~~P(1)]=1\label{CC11}\\
&&{\partial\over{\partial t_{\al,k}}}Q(1)=[Q(1),~~Q^k(\al)_-],\qquad \alpha
=1,2\label{CC12} 
\bj
In a similar way we can get the time evolution of $\Phi$ via a {\it discrete
linear system II}, whose consistency conditions are:
 
\ai
&&[\q(2),~~P(2)]=1,\label{CC21}\\
&&{\partial\over{\partial t_{\al,k}}}Q(2)=[Q^k(\al)_+,~~Q(2)]
\label{CC22} 
\bj
One can write down flows for $P(1)$ and $P(2)$ as well, but we will not need 
them in the sequel.

3) ${ {Reconstruction formulae}}$.
\noindent
The third element announced above is the link between the
quantities that appear in the linear system and in the quantum equations of
motion with the original partition function. We have
\a
{\d \over \d t_{\al, r}} \ln Z_N(t,c) = {\rm Tr} \Big(Q^r(\al)\Big), \quad\quad
\al = 1,2 \label{ddZ}
\b
It is evident that, by using the equations (\ref{CC12},\ref{CC22}) above 
we can express all the derivatives of $Z_N$ in terms of the elements of the 
$Q$ matrices. For example
\a
{\d^2\over{\d t_{1,1}\d t_{\al,r}}}
\ln Z_N(t,c)=\Bigl(Q^r(\al)\Bigl)_{N,N-1},\qquad \al = 1,2\label{parti3}
\b
and so on. We recall that the derivatives of $F(N,t,c) = \ln Z_N(t,c)$ at
prescribed values of the coupling are  
nothing but the correlation functions of the model.

The above derivation is rigorous when, for example, highest negative 
even couplings guarantee
that the measure in (\ref{orth1}) is square--integrable and decreases
more then polynomially at infinity. 
But for generic values of the couplings it is heuristic. 
Nevertheless we notice that the consistency and quantum equations of motion 
make sense for any value of the couplings, and also when the couplings are
infinite in number. In the latter case eqs.(\ref{CC12})
and eqs.(\ref{CC22}) form nothing but a very well--known
discrete integrable hierarchy, the Toda lattice hierarchy, see \cite{UT}. 

From these considerations it is clearly very convenient to refer to the 
integrable system formulation rather then to the original path integral 
formulation of our
problem. This allows us not only to extend our problem to a larger
region of the parameter space, but also to make full use of integrability.
 
To end this section, we collect a few formulas we will need later on.
First, we will be using the following coordinatization of 
the Jacobi matrices
\a
Q(1)=I_++\sum_i \sum_{l=0}^{m_1} a_l(i)E_{i,i-l}, \qquad\qquad\qquad
\q(2)=I_++\sum_i \sum_{l=0}^{m_2} b_l(i)E_{i,i-l}\label{jacobi1}
\b
where $I_+ = \sum_{i=0}E_{i,i+1}$ and $(E_{i,j})_{k,l}=
\delta_{i,k}\delta_{j,l}$. One can immediately see that
\a
\Bigl(Q_+(1)\Bigl)_{ij}=\delta_{j,i+1}+a_0(i)\delta_{i,j},\qquad
\Bigl(Q_-(2)\Bigl)_{ij}=R(i)\delta_{j,i-1}\label{jacobi2}
\b
where $R({i+1}) \equiv h_{i+1}/h_i$.
As a consequence of this coordinatization, eq.(\ref{parti3}) gives in
particular the two important relations
\a
{\d^2\over{\d t^2_{1,1}}}
F(N,t,c)=a_1(N),\label{Fa1}
\b
and
\a
\frac {\partial^2}{\d t_{1,1} \d t_{2,1}}F(N,t,c) = R(N)\label{FR}
\b
 
For reasons of brevity we do not even touch on the subject of W--constraints.
The latter are constraints on the partition function under the form of
algebraic structure, see \cite{BX1,BCX} for instance. They are obtained by
putting together quantum equations of motion and flow equations. 
W--constraints (which are also called loop equations or Schwinger--Dyson
equations) can be used to solve matrix models, but such a 
procedure is less efficient than the one used in the sequel.
 
\subsection{Homogeneity and genus expansion.}

The CF's we compute are genus expanded. The genus expansion is strictly
connected with the homogeneity properties of the CF's. 
The contribution
pertinent to any genus is a homogeneous function of the couplings (and $N$)
with respect to appropriate degrees assigned to all the involved quantities. 
Precisely, we assign to the couplings the following degrees
\a
deg(~) \equiv [~],\qquad [t_{\al,k}] = x (1-k),\qquad [N]=x,
\qquad [c] = -x \label{degree}
\b
where $x$ is an arbitrary positive number. Here and in the 
following $N$ is treated as a coupling $t_{1,0}=t_{2,0}$. 
If we rescale the couplings as follows
\a
t_{\al,k} \rightarrow \lambda^{[t_{\al,k}]} t_{\al,k}\0
\b
we expect the free energy to scale like
\a
F \rightarrow \sum_{h=0}^\infty \lambda^{x(2-2h)} F_h\label{genusexp}
\b
where $h$ is the genus. In other words
\a
\relax [F_h] = 2x(1-h) \label{degFh}
\b
$F_h$ is interpreted as the result of summing the open string partition function
at fixed $h$ over all the boundaries.

The CF's will be expanded accordingly.
Such expectation, based on a path integral analysis, remains true in our setup
due to the fact that the homogeneity properties carry over to the Toda lattice
hierarchy. To this end we have simply to consider a genus expansion for all 
the coordinate fields that appear in $Q(1)$ and $Q(2)$, see (\ref{jacobi1}, 
\ref{jacobi2}). 
The Toda lattice hierarchy splits accordingly.
In genus 0 the following assignments
\a
\relax [a_l^{(0)}] = (l+1)x,\qquad [b_l^{(0)}] = (l+1)x,\qquad 
[R^{(0)}] =2\, x \label{degToda}
\b
correspond exactly to the assignments (\ref{degree}) and $[F_0] = 2x$.

\section{Solving two--matrix models}

\setcounter{equation}{0}
\setcounter{footnote}{0}

As already pointed out in the previous section, a way to solve a two--matrix
model is to solve the coupling conditions (quantum equations of motion). 
This allows us to determine the 
`fields' $a_i(n), b_i(n)$ and $R(n)$. Once these are known we can compute 
all the correlation functions starting from (\ref{ddZ}) by repeated use of 
eqs.(\ref{CC11},\ref{CC12},\ref{CC21},\ref{CC22}), which form the flows of the
Toda lattice hierarchy. As for the free energy $F(t,N,c)$, see for 
instance (\cite{BX2}). In \cite{BCX}, using this method, the bi-Gaussian model 
${\cal M}_{2,2}$ was solved. This is of course a simple model. However it is 
useful to check the coincidence of the results obtained in this way in 
the decoupling $c=0$ case with the available results obtained by the 
traditional method based on eigenvalue density and resolvent for the 
Gaussian one--matrix model. 

A very interesting case is the model ${\cal M}_{0,0}$, i.e. the limiting
model when only the $c$ parameter is different from zero. As a path integral 
this model does not make much sense. However, as we saw above, it does have
sense as integrable system to which the appropriate coupling conditions are
applied. It turns out that this model describes $c=1$ string theory at the
self-dual radius, as was shown in \cite{BX2,BX3}.

\subsection{Solving the quantum EoM's:  ${\cal M}_{3,2}$ model}

The next model in order of complexity is the ${\cal M}_{3,2}$ model, \cite{BX3}.
The relevant quantum equations of motion are 
\a 
&& P^\circ(1) + 3t_3 Q(1)^2 + 2t_2 Q(1)  + t_1 + c Q(2) =0\label{cm321}\\
&& \widetilde {\cal P}^\circ(2) + 2s_2 Q(2) + s_1 + c Q(1) =0\label{cm322}
\b
Using the coordinatization (\ref{jacobi1}) and (\ref{jacobi2})
they produce the following equations for the fields $a_l(n), b_l(n), R(n)$ 
\a
&& cb_2(n) + 3t_3 R(n) R(n-1) =0 \0\\
&& 2t_2 R(n) + c b_1(n) + 3t_3 R(n)\Big (a_0(n) + a_0(n-1)\Big) =0\0\\
&& 3t_3 \Big( a_0(n)^2 + a_1(n) + a_1(n+1) \Big) 
+ 2t_2 a_0(n) +t_1 + c b_0(n) =0\0\\
&& n+ 3t_3 a_1(n) \Big( a_0(n) + a_0(n-1)\Big) + 2t_2 a_1(n) + c R(n) =0\0\\
&&2s_2 R(n) +ca_1(n)=0\label{coum32}\\
&& 2s_2 b_0(n) + s_1 + c a_0(n) =0 \0\\
&& n + 2 s_2 b_1(n) + c R(n) =0 \0
\b
where we have introduced the simplified notation
\a
t_{1,k} \equiv t_k,\qquad t_{2,k}\equiv s_k\0
\b
One easily realizes that the second, fourth, fifth and seventh equations are 
linearly dependent. The remaining equations determine the lattice fields
$a_0,a_1,b_0,b_1,b_2,R$ completely. The ${\cal M}_{3,2}$ model, even though 
comparatively simple is already rather complex due to the large number of
involved fields. Therefore, for pedagogical purposes, let us further simplify it, by
setting $c=0$ and considering only the $a_0,a_1$ fields. This corresponds to 
the one--matrix model with cubic interaction. In eqs.(\ref{coum32}), for
consistency, we have to set also $R=0$. The relevant equations are
\a
&& 3t_3 \Big( a_0(n)^2 + a_1(n) + a_1(n+1) \Big) 
+ 2t_2 a_0(n) +t_1 =0\0\\
&& n+ 3t_3 a_1(n) \Big( a_0(n) + a_0(n-1)\Big) + 2t_2 a_1(n)=0\label{1mcubic}
\b
One can derive $a_1$ from the second equation in terms of $a_0$ and replace 
it into the first. One gets in this way a cubic algebraic recursive equation
for $a_0$. We solve it with a genus by genus approach. The first step is to
start with genus 0. To do so one simply ignores the increments $\pm 1$ on the
$n$ entry. In this way we get an ordinary cubic algebraic equation in the
unknown $a_0$:
\a
&& a_0^3 + \frac {t_2}{t_3} a_0^2+ \frac 29 \left(\frac{t_2}{t_3}\right)^2 a_0  
- \frac n{3t_3}=0\label{cubica0}\\
&& a_1 = -\frac 12  a_0^2 - \frac 13 \frac {t_2}{t_3}  a_0\label{a1a0}
\b
where, for simplicity and without loss of generality, we have set $t_1=0$.
In the large $N$ limit, it is convenient to shift to the continuous formalism, 
by defining the continuous variable $x= \frac nN$. It is also useful
to make contact with section 4 of \cite{BIZ} for a comparison. So, also in order to 
simplify a bit further the notation, we set
$t_2=-\frac N2$ and $t_3=-Ng$, where $g$ is the cubic coupling constant there. 
Moreover we denote $f=3g a_0$. Then eq.(\ref{cubica0}) becomes
\a
18 g^2 x + f(1+f)(1+2f)=0\label{cubicf}
\b
It is easy to find the three solutions, which for small $x$ take the form
\a
&& f_1= -18 g^2 x - 972 g^4 x^2- 93312 g^6 x^3 - 11022480 g^8 x^4 +
O(x^5)\label{f1}\\
&& f_2= -1 -18g^2 x +972 g^4 x^2 -93312 g^6 x^3 + 11022480 g^8 x^4
+ O(x^5) \label{f2}\\
&& f_3 = -\frac 12 + 36 g^2 x + 186624 g^6 x^3 +O(x^5)\label{f3}
\b
From them we can easily write the fields $a_0$ and $a_1$ in terms of $g$ and $x$.
They therefore lead to three different solutions for the correlators.
Later on we will show how to compute the latter. But, before, let us discuss the 
meaning of these three solutions. To start with, comparing this with \cite{BIZ}, we see 
that the first solution corresponds to the unique solution found there provided 
we set $x=1$. It corresponds to the minimum of the classical potential.
In fact the correspondence with \cite{BIZ} can be made very precise: one can
easily verify that eqs.(46) there are nothing but 
eqs.(\ref{cubica0},\ref{a1a0}) provided we make the identifications:
$a+b=a_0$ and $(b-a)^2= 4 g a_1$ with $x=1$. 
In \cite{BIZ} the interval $(2a,2b)$ represents the cut in the eigenvalue
distribution function. This cut therefore has to be found in our formalism
in the $(a_0,a_1)$ plane, once we forget the dependence of the latter on $x$.

The classical potential for the continuous eigenvalue function $\lambda(x)$ 
(which is $\lambda_n/\sqrt{N}$ in the large $N$ limit), is 
$V_{cl}=\frac 12 \lambda^2 +g \lambda^3$. It has extrema at $\lambda =0$ and 
$\lambda= -1/3g$. To find the classical limit we have to drop the last term in
eq.(\ref{cubica0}). Remember that dropping the latter (with $c=0$)
is equivalent to writing 
eq.(\ref{cm321}) as $V_1'(Q(1))=0$. In other words the latter is the 
equation that identifies the extrema of the quantum potential $V_1$. They are 
three, $f=0,-1, -1/2$, which corresponds to $a_0=0, -1/3g, -1/6g$,  not two as 
in the classical case. So we see that $a_0$ approaches, in the classical limit, 
the classical eigenvalue function. Moreover
$f=0$ corresponds to the minimum of the potential, $f=-1$ corresponds to the 
maximum and $f=-1/2$ to the vanishing of the second derivative. The first 
two cases (the classical extrema) are characterized by the fact that $a_1=0$, 
while the third corresponds to a non-vanishing $a_1$.

From this simple example we learn three important pieces of informations.

\begin{itemize}  

\item The number of solutions of the quantum problem (i.e. the number of 
solutions to 
eq.(\ref{cubicf})) is larger than the number of the extrema of the classical 
potential.

\item The field $a_0$ can be regarded as the quantum version of the classical 
eigenvalue function.

\item The classical extrema are obtained by setting, together with $n=0$ in 
eq.(\ref{cubica0}), also $a_1=0$. 

\end{itemize}

As we shall see, the last condition, in the most general case, must be replaced 
by all fields $a_1,a_2,...$ present in the problem being set to zero (except
$a_0$).

What is the meaning of the third solution, $f=-1/2$? It is a non-perturbative
solution. It cannot be seen in the saddle point approximation.
More on this later on.

It is now easy to extend the analysis to the full ${\cal M}_{3,2}$ model. 
In genus 0 (i.e. disregarding the $\pm 1$ increments on $n$) it leads to 
the following set of equations, see \cite{BCX}:
\a
&&a_1(n) = - \frac {2s_2}{c} R(n), \qquad  
b_0(n) = - \frac{ c a_0(n)}{2 s_2}\0\\
&& b_1(n) = - \frac{n + c R(n) }{2 s_2}, \qquad   
b_2(n) = - \frac{3t_3}{c} R(n)^2\label{couplm32}
\b
and the recursion  relations
\a
&& 2a_0(n) = -\frac{2t_2}{3t_3} + \frac {c}{6s_2t_3} 
\Big( c + \frac{n}{R(n)}\Big)\label{recur1'}\\
&&2R(n) = \frac{c}{2s_2} a_0(n)^2 + \Big( \frac{2 c t_2}
{6s_2t_3} - \frac{c^3}{12 s_2^2 t_3}\Big) a_0(n)\label{recur2'}
\b
As expected the last two equations lead to the same cubic equation for $a_0$ 
as (\ref{cubicf}) with modified coefficients. It is interesting to find the 
classical extrema 
of this model. According to the above recipe we must set $a_1=b_1=b_2=0$, and
drop the first term in the third and sixth of eqs.(\ref{coum32}). 
In the genus 0 version of the latter this leads to $R=0$ and to
\a
&&3 t_3 a_0^2 +2t_2a_0 +cb_0=0\0\\
&& 2 s_2 b_0 +ca_0=0\label{vac32}
\b
The extrema correspond therefore to
\a
a=0,\quad\quad {\rm and}\quad\quad a_0 =\frac{c^2-4s_2t_2}{6s_2t_3}
 \label{class32}
\b
which is what one expects by completing the quadratures in the original 
classical potential. Of course, as above, in this way we find only two 
extrema out of three.

The solutions of the ${\cal M}_{3,2}$ model therefore can be found in the same
way as in the simplified one--matrix model above.

\subsection{Solving the quantum EoM's:  ${\cal M}_{4,2}$ model}

As a further example we briefly analyse the ${\cal M}_{4,2}$ model with, 
for simplicity, only the $t_4, t_2,c$ and $s_2$ couplings switched on. 
The quantum equations of motion are
\a 
&& P^\circ(1) + 4t_4 Q(1)^3 + 2t_2 Q(1)  + c Q(2) =0\label{cm421}\\
&& \widetilde {\cal P}^\circ(2) + 2s_2 Q(2) + c Q(1) =0\label{cm422}
\b
The matrices $Q(1)\in [-1,1]$ and $Q(2)\in [-1,3]$. 
Using the coordinatization (\ref{jacobi1}) and (\ref{jacobi2})
they produce the following equations for the fields $a_l(n), b_l(n), R(n)$, 
which we write down in the genus 0 version:
\a
&& n + 12 t_4(a_1^2+a_0^2 a_1) +2t_2 a_1+ cR=0\0\\
&&4t_4(a_0^3+6a_0a_1)+2t_2a_0+cb_0=0\0\\
&&12 t_4(a_0^2+a_1)+2t_2+c \frac{ b_1}{R}=0\0\\
&&12 t_4 a_0 +c\frac{b_2}{R^2}=0\0\\
&&4t_4 +c\frac{b_3}{R^3}=0\0\\
&&n +2s_2b_1+cR=0\0\\
&&2s_2b_0+ ca_0=0\0\\
&&2s_2R+ca_1=0\label{com42}
\b
Now let us proceed as in the ${\cal M}_{3,2}$ and set $c=0$. We obtain two 
decoupled one--matrix model systems, a Gaussian one on the right and a quartic 
one on the left. We are interested in the latter. The relevant equations are
\a
&& n + 12 t_4(a_1^2+a_0^2 a_1) +2t_2 a_1=0\0\\
&&4t_4(a_0^3+6a_0a_1)+2t_2a_0=0\label{coum4}
\b
Now, the second equation admit the solution $a_0=0$. Replacing it into the first
we obtain
\a
n + 12 t_4 a_1^2+ +2t_2 a_1=0\label{quadr}
\b
If $a_0\neq 0$, we can derive $a_1$ from the second equation and replace the 
result into the first, obtaining a biquadratic equation for $a_0$
\a
12 \frac {t_4}{t_2} \frac n{t_2}-20 \left( \frac {t_4}{t_2} \right)^2a_0^4 
-12  \frac {t_4}{t_2} a_0^2-1=0\label{biqua}
\b
Both (\ref{quadr}) and (\ref{biqua}) can be solved exactly. They give rise
to six distinct (in general complex) solutions. The classical potential for 
the eigenvalue function in the large $N$ limit is 
$V_4(\lambda)\sim t_2\lambda^2 +t_4\lambda^4$. This potential has one or 
three real solutions depending on whether
$t_2$ and $t_4$ have the same or opposite sign: $\lambda =0$ and 
$\lambda^2 = -\frac {t_2}{2t_4}$. In order to single out among the above six 
the solutions that correspond to those in the classical limit, we follow the 
above given recipe.
We drop the first term in the first eq.(\ref{coum4}) and set $a_1=0$ in 
both. We are left with 
\a
4t_4a_0^3+2t_2a_0=0\label{classm4}
\b
This gives exactly the expected classical extrema for $a_0$. Once these 
are determined we can easily find the corresponding quantum solutions either 
in exact form or in series of $x= n/N$. Following the example of the previous 
subsection we can also determine the 
solutions of the complete ${\cal M}_{4,2}$ model.

\subsection{Solving the quantum EoM's:  ${\cal M}_{3,3}$ model}

We study the model in the case $t_1=s_1=0$ and limit ourselves to writing down
the genus 0 quantum equations of motion:
\a
&&3t_3 c a_0^2 +2t_2 c a_0 -36 s_3t_3b_0R + c^2 b_0 -12s_2t_3R=0\0\\
&&3 s_3c b_0^2 +2s_2cb_0 -36s_3t_3a_0R -12s_3t_2R + a_0c^2=0\0\\
&&nc +Rc^2 -18 s_3t_3R^2 -36s_3t_3a_0b_0R -12 s_2t_3a_0R -12 t_2s_3b_0R
-4s_2t_2R=0\label{coum33}\\ 
&&a_1 = -\frac {6s_3}g b_0 R - \frac{2s_2}g R,\quad\quad\quad a_2= - \frac {3s_3}g
R^2\0\\
&&b_1 = -\frac {6t_3}g a_0 R - \frac{2t_2}g R,\quad\quad\quad b_2= - \frac {3t_3}g
R^2\0
\b
For simplicity we compute only the classical vacua. To this end
we drop the first term in the third
equation and solve the resulting system. Then we set
$a_2=b_2=a_1=b_1=0$ as well as $R=0$. In this branch we therefore have
\a
&& 3t_3 a_0^2+2t_2a_0 +cb_0=0\label{m33cl1}\\
&& 3s_3b_0^2+2s_2b_0 +c a_0=0\label{m33cl2}
\b
From the first we can get $b_0= -\frac 1c ( 3t_3 a_0^2+2t_2a_0 )$, whence
we get either $a_0=0$ or
the cubic equation
\a 
27 s_3t_3^2 a_0^3 + 36 t_2 s_3 t_3 a_0^2 +(12s_3t_2^2-6 c s_2t_3)a_0 +
c(c^2-4s_2t_2)=0\0
\b
Therefore in general we have four classical extrema, with non-vanishing
$a_0$ and $b_0$ while all the other fields vanish. A series expansion
about these solutions is easy to find. For instance around the vacuum 
$a_0=b_0=R=0$ we have
\a
&&a_0= \! \frac{ 12   ( -c   s_3  t_2 + 2  s_2^2  t_3)   x}{ ( c^2  
- 4  s_2  t_2)^2 } \0\\
&&~~~~~~~~~~+ \frac{648   [ -8   c  s_3^3  t_2^4  
+ c^2  s_3^2  t_2   (c^2 + 8  s_2  t_2)   t_3 - 
            4  c  s_2^2  s_3   (c^2 + 2  s_2  t_2)   t_3^2 + 
            16  s_2^5  t_3^3]   x^2}{(c^2 - 4  s_2  t_2)^5}  +{\cal O}(x^3)\0\\
&&b_0 = \!  -  \frac{12  ( -2  s_3 t_2^2 + 
                c s_2 t_3)  x }{ (c^2 - 4 s2 t2)^2} \0\\
		&&~~~~~~~~~~  + 
  \frac{648  [16 s_3^3 t_2^5 - 
            4 c s_3^2 t_2^2  (c^2 + 2 s_2 t_2)  t_3 + 
            c^2 s_2 s_3  (c^2 + 8 s_2 t_2)  t_3^2 - 
            8 c s_2^4 t_3^3]  x^2}{  (c^2 - 4 s_2 t_2)^5 } +{\cal O}(x^3)\0\\
&& R= \!  - \frac{ c x }{ c^2 - 
            4 s_2 t_2 }   +  \frac{18 c  [ -16  s_3^2 t_2^3 
	    + c s_3  (c
^2 + 12 s_2 t_2)  t_3 - 16 s_2^3 t_3^2]  x^2 }{ (c^2 - 
4 s_2 t_2)^4 }+{\cal O}(x^3)\0
\b	    
For reasons of space we have limited the expansion in $x=n/N$ to the quadratic
order. From (\ref{coum33}) one can easily compute the expansions for
$a_1,b_1,a_2,b_2$.

\subsection{The ${\cal M}_{p_1,p_2}$ model}

In the general case the matrix rank for $Q(1)$ and $Q(2)$ was given in 
the previous section and the quantum EoMs become of course very complicated.
It is however simple to write down the equations that identify the 
extrema with classical analog. They are
\a
V_1'(a_0) + c b_0=0,\quad\quad\quad V'_2(b_0) +c a_0=0\label{extremaeq}
\b
while all the other fields are set to zero. We have in general 
$(p_1-1)(p_2-1)$ solutions of this type in perfect correspondence with the
classical analysis. The simplest solution is $a_0=b_0=0$. Other solutions may 
be hard or even impossible to determine explicitly. Anyhow, once one such
solution is known it is possible to find explicit expressions for the fields
around it in terms of $x=n/N$.

\subsection{Calculating the correlators}

Once we know the fields $a_i,b_j,R$ in a given model there exists an 
algorithmic procedure to determine the correlators. This in turn is due to the
integrability underlying the Toda lattice hierarchy. In this subsection we
give a few examples of exact correlators. The general scheme is known, it was 
already presented in \cite{BX1,BCX}. A few explicit examples were worked out 
for the ${\cal M}_{0,0}$ model in \cite{BX2,BX3}. In these references one 
can find explicit calculations of correlators for finite $N$ and for any genus. 
In this subsection we limit ourselves to large $N$ 
genus 0 correlators. To start with let us briefly review the continuous 
versions of the quantum equations of motion and the flow equations in this 
case (which are known as the dispersionless Toda lattice hierarchy flows).

We proceed as in \cite{BX1}. First we define the continuum quantities
in the following way  
\def\rt{ t^{\rm ren} }
\def\rs{ s^{\rm ren} }
\a
x= \frac{n}{N}, \qquad t^{\rm ren}_k = \frac{t_k }{N},
\qquad \rs_k = \frac{s_k }{N}, \qquad c^{\rm ren}= \frac{c}{N} \0
\b
in the large $N$ limit. They are the {\it renormalized} coupling constants.
In the following however we will understand the superscript ${\rm ren}$. 
Further we define
\a
F_0(x) = \lim_{N\to\infty}  \frac{F_N}{N^2},\qquad
\zeta = \lim_{N\to\infty}  I_+ \0
\b
where $F_0$ is the genus zero free energy. The second equation is merely
symbolic and simply means that $\zeta$ is
the continuum counterpart of $I_+$.
If we define a matrix $\rho = \sum_n n E_{n,n}$, it is easy to see that we have
\a
\relax [ I_+, \rho] = I_+, \label{FPd}
\b

The continuous counterpart gives the following basic Poisson bracket
\a
\{ \zeta , x \} = \zeta. \label{FP}
\b
This allows us to establish the following correspondence
\a
N~[~~,~~] \Longrightarrow \{~~,~~\} \label{ctp}
\b
and, similarly,
\a
\frac{1}{N} {\rm Tr} = \frac{1}{N} \sum_{0}^{N-1} \Longrightarrow \int_0^x dx
 \label{ctrint}
\b
together with the replacements
\a
Q(1) \rightarrow L, \qquad Q(2) \rightarrow \tl; \0 
\b
where
\a
L=\zeta + \sum_{l=0}^\infty a_l \zeta^{-l}, \qquad
\tl=\frac{R}{\zeta} + \sum_{l=0}^\infty \frac{b_l}{R^l} \zeta^l. \label{ltl}
\b
Here $a_l$ and $b_l$ are the continuum fields that replace the lattice fields
$a_l(i)$ and $b_l(i)$ of eq.(\ref{jacobi1}).

We stress that the above replacements holds only in genus 0.

Now let us come to the flow equations:
\noindent{\it The dispersionless limit of the extended 2-dimensional Toda
lattice integrable hierarchy (\ref{CC12}, \ref{CC22}) is}
\ai
&&{{\d L}\over {\d t_{k}}} = \{ L, ~~ (L^k)_-\}, 
\qquad\quad{{\d L}\over {\d s_{k}}}=\{L,(\tl^k)_-\},\label{distoda1}\\
&&{{\d \tl}\over {\d t_{k}}} = \{ (L^k)_+, ~~\tl\},
\qquad\quad{{\d \tl}\over {\d s_{k}}} = \{ (\tl^k)_+, ~~\tl\} \label{distoda2}
\bj
Here the subscript + denotes the part containing non--negative
powers of  $\zeta$, while -- indicates the complementary part.
 
Next, the continuum version of (\ref{ddZ}) provides
{\it the link between the free energy and Lax operators}, i.e.
\a
\frac{\d}{\d t_{k}}F = \int_0^x\, (L^k)_{(0)}(y)dy 
\qquad \frac{\d}{\d s_{k}}F = \int_0^x\, (\tl^k)_{(0)}(y)dy \label{ddf}
\b
where subscript `${}_{(0)}$' means that we select the coefficient of the 
zero--th power term of $\zeta$.  This formula opens
the way to calculate CF's by simply differentiating both sides
with respect to the appropriate coupling constants and use eqs.(\ref{distoda1},
\ref{distoda2}). Therefore this equation
together with the integrable hierarchy and the quantum equations of motion
completely determines the genus zero correlators.  
For instance
\a
<\tau_k \tau_l> =\frac{\d^2 F} {\d t_{k} \d t_{l} } = \oint
 (L^l)_- d L^k,\qquad 
<\tau_k \sigma_l> =\frac{\d^2 F} {\d t_{k} \d s_{l} } = \oint
 (\tl^l)_- d L^k, 
 \label{h2pf}
\b
and
\a
 <\sigma_k\sigma_l>=\frac{\d^2 F}{\d s_{k} \d s_{l} }=-\oint
(\tl^l)_{\geq1}d\tl^k. \label{h2pf'}
\b
where $\oint$ represents the residue at the simple pole in $\zeta$.

It is now very easy to extract explicit expressions for correlators in various
model. Here we limit ourselves for simplicity to a simple example,
the ${\cal M}_{3,2}$ model
in the decoupling limit studied in subsection 2.1. In this case we have two
fields $a_0$ and $a_1$ and explicit expressions for them. One starts
from eqs.(\ref{f1},\ref{f2},\ref{f3}), then computes $a_0= -\frac{f}{3g}$ and 
finally $a_1$ from eq.(\ref{a1a0}). The Lax operator is given in this case by
\a
L= \zeta+ a_0+\frac {a_1}\zeta\label{L}
\b
Inserting it in the previous formulas we get
\a
<\tau_k> = \sum_{2l=0}^k \frac {k!}{(k-2l)! l!l!} \int_0^x a_0(y)^{k-2l} a_1(y)^l
dy\label{tauk}
\b
and
\a
&&<\tau_k\tau_r>= \label{tauktaul}\\
&&\sum_{l=0}^r \sum_{p=0}^{[\frac {l-1}2]} \sum_{n=0}^{k-1}
\sum_{q=0}^n \frac {k!}{(k-1-n)!(n-q)!q!} a_0^{r-l+k-n-1} a_1^{l-p+n-q}
\left[\delta_{2p+2q-l-n,-1}-a_1\delta_{2p+2q-l-n,1}\right]\0
\b  
where $[\alpha]$ denotes the integral part of $\alpha$.  
Replacing eqs.(\ref{f1},\ref{f2},\ref{f3}), as indicated above, we find
explicit expressions for the correlators in terms of $x$ and $g$.
 
Similarly one can compute the three point functions (referred to generically as
Yukawa couplings) 
\a
<\tau_{l_1}\tau_{l_2}\tau_{l_3}>\!&&\!=~~~ \prod_{i=1}^3\left(\sum_{k_i=0}^{l_i}
\sum_{p_i=0}^{k_i} \frac {{l_i}!}{(l_i-k_i)!(k_i-p_i)!p_i!}\right)
\,\delta_{\sum_{i=1}^3 k_i\, , \, 2\sum_{i=1}^3 p_i}\cdot\label{threep}\\
&&\left[a_0^{l_1+l_2 -k_1-k_2}a_1^{p_1+p_2} \frac {d}{dx} 
\left(a_0^{l_3-k_3}a_1^{p_3}\right) (k_1-2p_1)(k_2-2p_2)\,\theta(k_2-2p_2)\,
\theta(2p_3-k_3)\right.\0\\
&&-a_0^{l_1+l_3 -k_1-k_3}a_1^{p_1+p_3} \frac {d}{dx} 
\left(a_0^{l_2-k_2}a_1^{p_2}\right) (k_1-2p_1)(k_3-2p_3)\,\theta(k_1-2p_1)\,
\theta(2p_3-k_3)\0\\
&&\left. +a_0^{l_2+l_3 -k_2-k_3}a_1^{p_2+p_3} \frac {d}{dx} 
\left(a_0^{l_1-k_1}a_1^{p_1}\right) (k_2-2p_2)(k_3-2p_3)\,\theta(2p_2-k_2)\,
\theta(2p_3-k_3)\right]\0
\b
where $\theta(x)$ means 0 for $x\leq 0$ and 1 otherwise.

\subsection{Higher genus}

We would like to introduce in this subsection a few basic notions concerning
higher genus correlators. They are introduced here in order to render the
subsequent discussion as self--contained as possible. For a more complete
treatment see \cite{BX2,BX3,BCX}.

Let us consider the first example above, the decoupled ${\cal M}_{3,2}$ model,
whose general solutions are characterized by eq.(\ref{1mcubic}) and genus 0 
ones are explicitly given in eqs.(\ref{f1},\ref{f2},\ref{f3}).
Let us start from a given genus 0 solution, specified by $a_0=r_0$ and
$a_1=s_0$, where $r_0/3g$ is anyone of the three solutions 
(\ref{f1},\ref{f2},\ref{f3}). Then we expand the full solution in series of 
$\e=1/N$ as follows:
\a
a_0= \sum_{n=0} \e^n r_n(x),\quad\qquad a_1= \sum_{n=0} \e^n s_n(x) \label{expeps}
\b
Moreover $f(n\pm 1)$ is replaced by $ e^{\pm\e\d_x} f(x)$  for any lattice 
function $f(n)$. Inserting this into (\ref{1mcubic}) we obtain the genus 0
equations and an infinite series of relations for the higher order terms, which
can be recursively solved. For instance, the next to leading equations are
\a
&&6t_3 r_0r_1 + 3 t_3(s_0' + 2s_1) + 2 t_2 r_1=0\0\\
&& 6t_3 (s_1 r_0 +s_0r_1)- 3t_3 s_0r_0' + 2t_2 s_1=0\label{firstorder}
\b
where a prime denotes derivative with respect to $x$.
They can be easily solved and lead to
\a
&&r_1 = - \frac 32 t_3 \frac{3t_3 s_0r_0' + t_2 s_0' +3t_3 r_0s_0'}{t_2^2 +
6t_2t_3r_0+ 9t_3^2(r_0^2-s_0)}\0\\
&& s_1= \frac 32 t_3 s_0\frac{3t_3(s_0'+r_0r_0')+t_2r_0'}{t_2^2 +6r_0t_2t_3
+9(r_0^2-  s_0)t_3^2}\0
\b
Similarly, for the second order we get
\a
&& r_2= -\frac 34 t_3 \frac{t_2(2 r_1^2 +s_0''+2 s_1')+ 3t_3 (2r_1' s_0-r_0'' s_0
+ 2 r_0' s_1 - 4 r_1s_1 +r_0 (2r_1^2 +s_0^2+2 s_1'))}{t_2^2 +6
r_0t_2t_3+9t_3^2(r_0^2-s_0)}\0\\
&& s_2 = \frac 34 t_3 \frac{- 3s_0 t_3 (2r_1^2 + s_0^2 + 2s_1') + (t_2+3
r_0t_3)(r_0'' s_0 - 2 r_1' s_0 - 2r_0' s_1+4 r_1 s_1)}{9 s_0t_3^2 -(t_2+3
r_0t_3)^2}\0
\b 
and so on. Replacing these expressions into the appropriate formulas
for the correlators, we can write down their explicit genus expansions.
To this end one should recall that the appropriate expansion for 
correlators is given by a power series in $\e$, as one can infer from
the free energy expansion
\a
F(x,\e) = \sum_{h=0}^\infty F_h(x) \e^{2h}\label{Fexpan}
\b
where $h$ is the genus.

We give, as an example, the genus 1 contribution to $<\tau_k>$, which is
the coefficient of $\e^2$ in the $\e$ expansion:
\a
<\tau_k>_1\!\!\!\!&&=\int_0^xdy \left[ 3 {n \choose 4} \left( A^{n-4}_{(-2)}
\, r_0'^2 +2 A^{n-4}_{(-1)}\,r_0's_0' +A^{n-4}_{(0)} (s_0'^2
-2r_0r_0'^2)\right.\right.\0\\
&& \left.-4 A^{n-4}_{(1)} \, s_0r_0's_0' + A^{n-4}_{(2)} (s_0^2r_0'^2- 2s_0s_0'^2)+
2A^{n-4}_{(3)} s_0^2r_0's_0' + A^{n-4}_{(4)}\ s_0^2s_0'^2\right)\0\\
&& +{n \choose 3} \left( A^{n-3}_{(-1)}r_0'' +  A^{n-3}_{(0)}(r_0'^2 + s_0''+3
r_0' r_1) \right.\0\\
&&\left.+  A^{n-3}_{(1)} (2 r_0's_0'-s_0r_0''-s_0'r_0'+3r_0's_1+3 s_0' r_1)
 +  A^{n-3}_{(2)}(s_0'^2-s_0s_0''-s_0'^2+3 s_0's_1)\right)\0\\
&& +{n \choose 2}\left( A^{n-2}_{(-2)}r_1' + A^{n-2}_{(-1)}(\frac 12 r_0'' 
+ s_1')+ A^{n-2}_{(0)}(r_1^2-2 s_0r_1') \right.\0\\
&&\left.
+ A^{n-2}_{(1)}(\frac 12 s_0 r_0'' + \frac 12
s_0''-s_1 r_0' - 2s_0s_1'+ 2 r_1 s_1)
+ A^{n-2}_{(2)} (r_1's_0^2 +s_1^2-s_1s_0')+
A^{n-2}_{(3)}s_0^2s_1'\right)\0\\
&& + \left.{n \choose 1} \left( A^{n-1}_{(0)}r_2 + 
A^{n-1}_{(1)}s_2\right)\right]\label{tauk'}
\b  
where $r_0,s_0,r_1,s_1,r_2,s_2$ are function of $y$ and
a prime denotes differentiation with respect to $y$. Moreover
\a
A^{n}_{(k)}= \sum_{2p+k=0}^{n}\frac {n!}{(n-2p-k)!(p+k)!p!} r_0^{n-2p-k}s_0^p
\label{Ank}
\b

\section{Two--matrix models and multiple brane configurations}

All the examples of the previous sections represent, according to the 
geometric description of section 2, the physics of $N$ D--branes wrapped 
around the two--dimensional sphere located in one of the vacua. 
A related problem is to describe a more complex situation with $N_1$ D--branes
at one vacuum and $N_2=N-N_1$ at another. There may of course be even more
complicated configurations with several groups of D--branes in different vacua.
Let us call them {\it multiple brane configurations}. However the example with
two groups of D--branes 
will be sufficient to illustrate the salient features of the problem.  
Let us consider once again the ${\cal M}_{3,2}$ model in the decoupling limit
so that we can work with explicit formulas. We refer in particular to
eq.(\ref{cubicf}), which we rewrite here in the form
\a
18 g^2 x+z(1+z)(1+2z)=0\label{cubicz}
\b
This can be solved exactly for $z$ and gives the three solutions
\a
&&z_1= -\frac 12 + \frac 1{2 I(x)} + \frac {I(x)}6\label{z1}\\
&&z_2= -\frac 12 + \frac {1+i\sqrt{3}}{4 I(x)} + \frac {1-i\sqrt{3}I(x)}{12}
\label{z2}\\
&&z_3= -\frac 12 + \frac {1-i\sqrt{3}}{4 I(x)} + \frac {1+i\sqrt{3}I(x)}{12}
\label{z3}
\b
where
\a
I(x)= 3^{1/3}\left(-324 g^2 x +\sqrt{3} \sqrt{-1+34992 g^4 x^2}\right)^{1/3}
\label{I(x)}
\b
When expanded for small $x$ they give rise to the series
(\ref{f1},\ref{f2},\ref{f3}), respectively. However this is not a very 
illuminating way of studying eq.(\ref{cubicz}). The best way is to consider
it a plane curve, \cite{BK}, in the complex $z,x$ plane. Then it represents
a genus 0 Riemann surface with punctures at $x=0$ and $x=\infty$. It is made of 
three sheets joined through cuts running from $z=- 1/(\sqrt{3} 108 g^2)$
to $z= 1/(108\sqrt{3}g^2)$.
The solutions (\ref{f1},\ref{f2},\ref{f3}) correspond to the values $z$ takes
near $x=0$, away from the cuts. Therefore we can pass from one solution to
another by crossing the cuts. We call the Riemann surface so constructed
the {\it quantum Riemann surface associated to the model}.
This Riemann surface picture is the clue to 
understanding the solutions with multiple brane configurations. Let us see how.

Let us start from ${\cal M}_{3,2}$, set $c=0$ and concentrate on the cubic 
interaction part. The relevant equations are (setting $t_1=0$)
\a
&& 3t_3 \Big( a_0(n)^2 + a_1(n) + a_1(n+1) \Big) 
+ 2t_2 a_0(n) =0\0\\
&& n+ 3t_3 a_1(n) \Big( a_0(n) + a_0(n-1) \Big) + 2t_2 a_1(n)=0\label{discubic}
\b
This equation in genus 0 has three solutions. Let us denote by $a_0=r_0$ one
such genus 0 solution and $a_1=s_0$ the corresponding $a_1$. Similarly, we pick
another solution and we denote it $a_0=u_0$ and $a_1=v_0$. They represent the
lowest order contribution of expansions like those considered in section 2.6.
\a
&&R(n) \to R(x) = \sum_{n=0}^\infty \e^n r_n(x)\0\\
&&S(n)  \to S(x) = \sum_{n=0}^\infty \e^n s_n(x)\label{RS}\\
&& U(n)  \to U(x) = \sum_{n=0}^\infty \e^n u_n(x)\0\\
&&V(n) \to V(x) = \sum_{n=0}^\infty \e^n v_n(x)\label{UV}
\b
where we have indicated the large $N$ expansion.
$R(n), S(n)$ and $U(n),V(n)$ form, separately, two couples of solutions of 
(\ref{discubic}). We recall that in the analogous problem formulated in the
familiar saddle point approach one sets $N_1$ eigenvalues $\lambda_i$ in 
one vacuum and $N_2$ in another. Considering the analogy between the field $a_0$ in
genus 0 and the classical eigenvalues, we are led
to pose the following problem: does there
exist a solution of (\ref{discubic}) that corresponds to $R(n), S(n)$ for 
$0\leq n\leq N_1-1$ and to $U(n),V(n)$ for $N_1\leq n\leq N-1$? 
This means that (\ref{discubic}) must hold with $a_0(n), a_1(n)$ replaced by
$R(n),S(n)$ for $0\leq n\leq N_1-1$ and   by $U(n),V(n)$ for 
$n\geq N_1$, respectively. However in addition we have the boundary equations
\a
&& 3t_3 \Big( R(N_1-1)^2 + S(N_1-1) + V(N_1) \Big) 
+ 2t_2 R(N_1-1) =0\0\\
&& N_1+ 3t_3 V(N_1) \Big( U(N_1) + R(N_1-1)\Big) + 2t_2 V(N_1)=0\label{boundary}
\b
While all the other equations are the same as in  
the previous section, these two equations represent the real novelty: they mix
two different solutions. They are
obviously satisfied if it makes sense to identify
\a
V(N_1) = S(N_1),\quad\quad\quad R(N_1-1)=U(N_1-1)\label{boundarycon}
\b
In the discrete formalism this is not easy to check. Therefore we shift to
the continuous formalism. Recalling what we did in subsection 2.5, 2.6, 
in the large $N$ limit we set
$N_2= \alpha N_1$ , $N=(1+\alpha)N_1$, $N_1/N=1/(1+\alpha) =\beta$. We also 
define $n/N =x$ for $0\leq n\leq N_1-1$ and 
$n/N = N_1/N + (n-N_1)/N =\beta +y$. 
In this formalism eqs.(\ref{boundarycon}) read
\a
V(\beta)= \lim_{\e\to 0} S(\beta+\e),\quad\quad R(\beta) =\lim_{\e\to 0}
U({\beta-\e})\label{contbound}
\b
This is nothing but the statement that at $x=\beta$ we are crossing the cut 
that separates the two solutions. In hindsight this is quite obvious: 
the only way to satisfy eq.(\ref{discubic}) with two different solutions 
is to cross the cut that join the corresponding sheets in the Riemann
surface introduced above. 

With this recognition in mind one can now set out to calculate correlators
in a theory with two sets of $N_1$ and $N_2$ D-branes in two different vacua.
Leaving a more complete treatment for another occasion we can easily
exhibit as an example the two point correlators analogous to (\ref{tauktaul}).
The correlator is given by the sum of two terms, each one equal to the right 
hand side of (\ref{tauktaul}): in the first $a_0,a_1$ are replaced by 
$r_0,s_0$ evaluated in $x$, while in the second they are replaced by
$u_0,v_0$ evaluated in $\beta+y$.

The conclusion is therefore that quantum solutions to (\ref{1mcubic})
that correspond to two groups of D--branes do exist. Moreover, due to the
structure of the quantum EoMs which we have explored in the previous  
section, it is easy to generalize this conclusion. Every model will be
characterized by a quantum Riemann surface with cuts that separate different
solutions. It is therefore possible to construct multiple brane 
configurations, as quantum solutions of the matrix model, by means 
of solutions of the quantum EoMs on different sheets that join to one another
across the appropriate cuts.

\section{Conclusions}

In this paper we considered B-model D-branes on 2-cycles of local 
Calabi--Yau geometries. The theory describing these objects is given by the 
reduction to the D-brane world-volume of the open string theory with Dirichlet 
boundary conditions on it.
We have described a precise dimensional reduction scheme for the 
holomorphic Chern-Simons
theory, that is the B-model open string field theory, to the 2-cycle.
This has been done for generic local CY geometries modeled around an 
arbitrary Riemann surface.
In the case of the conifold geometry, i.e. when the 2-cycle is a ${\mathbb P}^1$,
we have considered the relevant effective theory and found that it is given 
by a multi-matrix model.
The number of matrices involved depends on the reference complex structure 
about which we calculate the coupling to the Calabi--Yau complex moduli. 
The multi-matrix potential
is fixed by the complex moduli in a well defined and simple way.
The various allowed couplings turn out to be in correspondence with the projective 
parameters of complex structure deformation.
The matrix models we have obtained are of generic type if they involve one or 
two matrices, while we found relevant constraints within their 
parameters for more than two matrices.

We have studied the geometric engineering of the multi-matrix models 
and provided both a general reduction scheme and some examples.
Actually, some multi-matrix models are reducible to models involving less 
matrices. This phenomenon has a clear counterpart on the geometrical side, 
corresponding to the fact that different reference conifold complex 
structures can be connected via specific deformations.

In the second part of the paper we have focused 
on two-matrix models with bilinear couplings. We have illustrated
a general method to {\it exactly} solve these models. It consists in solving 
the quantum equations of motion
and making subsequently use of the integrable flow equations.
We have exhibited several examples of solutions in genus 0 as well as 
in higher order approximations. We have also discussed the relation of our 
method to the more popular saddle point approximation. One of the relevant
differences is that our method leads in general to more solutions than the
saddle point one. 

Finally we have discussed the brane interpretation of
our results. They represent the deformations of the complex structure
generated by the strings attached to N D--branes wrapped around a 2-cycle
in a local Calabi-Yau geometry. However it is possible to obtain more general
solutions: group of D-branes localized near different vacua. This is due 
to a remarkable property of our approach: all the data of all the solutions
of a given model are encoded in a Riemann surface (a plane curve), which we 
call the {\it quantum Riemann surface} of the model; different solutions  
lie on different sheets; there is room for solutions representing D-branes
localized near different vacua by crossing the cuts that connect the sheets.

To conclude the paper we would like to list a few open questions.
The first concerns geometric transitions and gauge duals. 
Laufer's theorem, \cite{laufer}, implies that only few smooth geometries 
can be interpreted as resolution of the singular conifold. 
These, as already observed in \cite{Ferrari}, correspond to 
asymptotically free gauge theories duals.
Then, for the non Laufer's geometries one should formulate a definite gauge dual.
Among these, one should find the non asymptotically free theories.
It seems natural to guess the gauge dual of the geometries ${\cal O}(-n)\oplus{\cal O}(n-2)$
to be then given by ${\cal N}=1$
SYM with $n+1$ Wess-Zumino multiplets in the adjoint representation of the 
$U(N)$ gauge group and with superpotential given by the 
corresponding multi-matrix model one.
Unfortunatly, we don't have convincing arguments to push further 
this hypothesis.

Secondly, we elaborated a scheme which can be applied to Calabi-Yau manifolds 
of more 
general type than conifolds. It would be interesting to work out 
the effective coupling to the CY complex moduli of the reduced theory in such 
more general settings.

The last comment concerns multi--matrix models. It is apparent that we have a 
method to solve any kind of two-- or multi--matrix model with bilinear 
couplings. The next important step is to find analogous powerful tools 
to solve matrix models with more complicated couplings.

\vskip 1cm

{\bf Acknowledgments.}  
We thank
M.~Bertolini,R.~Dijkgraaf,
T.~Grava,M.~Mari\~no,
L.~Mazzucato,
M.~S.~Narasimhan,
A.~Tanzini
for stimulating discussions.

This research was supported by the Italian MIUR
under the program ``Teoria dei Campi, Superstringhe e Gravit\`a''.
The work of G.B. is supported by
the Marie Curie European Reintegration Grant MERG-CT-2004-516466,
the European Commission RTN Program MRTN-CT-2004-005104.

\vskip 1cm

\section{Appendix}

In this appendix we extend the method for the reduction of the holomorphic CS 
functional to the non compact CY geometry around a four cycle.
This geometry, once one fixes the complex structure on the four manifold $M$ 
describing the cycle, is 
fully determined to be total space of the canonical line bundle $K_M$, 
that is the bundle of 
the top holomorphic forms on $M$. We denote this space as $X_M=tot(K_M)$.

Any atlas $\{ U_{(\alpha)} \}$
on $M$ extends to an atlas on $X_M$ 
by $\hat U_{(\alpha)} = U_{(\alpha)}\times {\mathbb C}$.
The complex manifold is defined by the patching conditions 
$$
{\bf z}_{(\alpha)}={\bf f}_{(\alpha)(\beta)}\left({\bf z}_{(\beta)}\right) \\
$$
\begin{equation}
p_{(\alpha)}= [\det {\bf X}_{(\alpha)(\beta)}]^{-1}
p_{(\beta)},\quad{\rm where}\quad
[{\bf X}_{(\alpha)(\beta)}]=
\partial_{{\bf z}_{(\beta)}}{\bf f}_{(\alpha)(\beta)}
\label{linear'}\end{equation}
in any double patch intersection $U_{(\alpha)} \cap U_{(\beta)}$.
In (\ref{linear'}) and in the following, $\z=(z^1,z^2)$ denotes the two 
complex coordinates. The holomorphic $(3,0)$-form on $X_M$ is 
$\Omega=dz^1\wedge dz^2\wedge dp$.

Let us consider the topological B-model on 
$X_M$. In this case, D-branes can wrap the 4-cycle $M$ and
the theory describing the dynamics of these objects is obtained then by 
reducing the HCS functional to the D-brane world-volume.
We consider here again only the case in which the gauge bundle $E$ is trivial.

The action of hCS is
\begin{equation}
S(\A)=\frac{1}{g_s}\int_{X_M} {\cal L}, 
\quad
{\cal L}=
\Omega\wedge Tr\left(\frac{1}{2} \A\wedge
\bar\partial \A + \frac{1}{3} \A \wedge \A \wedge \A\right)
\label{hCS'}\end{equation}
where $\A\in T^{(0,1)}\left(X_M\right)$.

We split $\A=\A_{\bar\z}d\bar\z +\A_{\bar p}d\bar p$
and we set, because of the glueing prescriptions for the parallel and transverse components,
$\A_{\bar\z}d\bar\z=A_{\bar\z}d\bar\z-A_{\bar p}\Gamma_{\bar\z}\bar p d\bar\z$
and $\A_{\bar p}=A_{\bar p}$, where 
$A=A_{\bar\z}d\bar\z\in T^{(0,1)}\left(M\right)$
is an anti-holomorphic 1-form on $M$, $A_{\bar p}\in\Gamma(\bar K_M^{-1})$
a section of the inverse anti-canonical line bundle and $\Gamma_{\bar\z}d\z$
is the $(0,1)$ component of a reference connection on $\bar K_M$.

The reduction prescription is that the matrix valued dynamical fields 
$(A_{\bar\z},A_{\bar p})$
are independent on the coordinate $p$ along the fibre ${\mathbb C}$.

Direct calculation of the Lagrangian ${\cal L}$ in (\ref{hCS'})
for the above reduced configurations gives 
\begin{equation}
L={\cal L}_{red}=\Omega\wedge Tr
\left(\frac{1}{2}
\left\{A\wedge\bar D A_{\bar p} +A_{\bar p} F^{(0,2)}+A\wedge\Gamma A_{\bar p}
\right\}
\right)d\bar p
\label{aredlag}\end{equation}
where $\bar D$ is the covariant derivative w.r.t. $A$. Notice that the above 
result does not depend on $p$.

Introducing now a reference section $K\in\Gamma(K_M\otimes \bar K_M)$, we 
define $\phi^{(2,0)}=K A_{\bar p}\in\Gamma(K_M)$
and fix the reference connection to be $\Gamma=K^{-1}\bar\partial K$. This 
way we get
\begin{equation}
L={\cal L}_{red}=\Omega K^{-1}\wedge Tr
\left(\frac{1}{2}
\left\{A\wedge\bar D \phi^{(2,0)} +\phi^{(2,0)} F^{(0,2)}
\right\}
\right)d\bar p
\label{aredlag'}\end{equation}

To reduce to a 4-form, we saturate the reduced Lagrangian with 
$K\partial_p\wedge\partial_{\bar p}$
so that the reduced HCS functional becomes just
\begin{equation}
HCS_{red}=\frac{1}{g_s}\int_M Tr
\left(\frac{1}{2}
\left\{A\wedge\bar D \phi^{(2,0)} +\phi^{(2,0)} F^{(0,2)}
\right\}
\right)=
\frac{1}{g_s}\int_M Tr\left(
\phi^{(2,0)} F^{(0,2)}
\right)
\label{aredlag''}\end{equation}
which is the form used in \cite{Diaconescu}.

\end{document}